\documentclass[letterpaper, 10 pt, conference]{ieeeconf}  % Comment this line out if you need a4paper

\let\proof\relax 
\let\endproof\relax 
\usepackage{booktabs}
\pdfoutput=1
\usepackage{bbm}
\usepackage{mathptmx}
\usepackage{graphicx} % more modern
\usepackage{amssymb}
\usepackage{amsthm,amsfonts,bm}
\usepackage{amsmath,mathtools}
\usepackage{graphicx} % more modern
\usepackage{caption,cite}
\usepackage{tikz}
\usetikzlibrary{arrows}

\usepackage{subfig} 
\newtheorem{definition}{Definition}
\newtheorem{lemma}{Lemma}
\newtheorem{assumption}{Assumption}
\newtheorem*{assumption*}{Assumption}
\newtheorem{proposition}{Proposition}

\newtheorem{corollary}{Corollary}

\IEEEoverridecommandlockouts                              % This command is only needed if 
\title{\LARGE \bf Matching Protocol and Pricing for Large-Scale Data Sharing in Massive IoT Networks
}

\author{Tao Zhang, Quanyan Zhu% <-this % stops a space
\thanks{}% <-this % stops a space
\thanks{Tao Zhang and Quanyan Zhu are with Department of Electrical and Computer Engineering, New York University, 370 Jay Street, Brooklyn, 11201, NY, USA.
        {\tt\small {\{tz636, qz494\}}@nyu.edu}}%
}

\begin{document}

\maketitle
\thispagestyle{empty}
\pagestyle{empty}

%%%%%%%%%%%%%%%%%%%%%%%%%%%%%%%%%%%%%%%%%%%%%%%%%%%%%%%%%%%%%%%%%%%%%%%%%%%%%%%%
%%%%%%%%%%%%%%%%%%%%%%%%%%%%%%%%%%%%%%%%%%%%%%%%%%%%%%%%%%%%%%%%%%%%%%%%%%%%%%%%

\begin{abstract}
The development of the Internet of Things (IoT) generates a significant amount of data that contains valuable knowledge for system operations and business opportunities. Since the data is the property of the IoT data owners, the access to the data requires permission from the data owners, which gives rise to a potential market opportunity for the IoT data sharing and trading to create economic values and market opportunities for both data owners and buyers. In this work, we leverage optimal mechanism design theory to develop a monopolist matching platform for data trading over massive IoT networks. 
The proposed mechanism is composed of a pair of matching and payment rules for each side of the market.
%
%The proposed mechanism is composed of a pair of matching and payment rules for each side of the market to optimize the payoffs under welfare-maximization and the revenue maximization schemes.
%
We analyze the incentive compatibility of the market and characterize the optimal mechanism with a class of cut-off matching rules for both welfare-maximization and revenue-maximization mechanisms and study three matching behaviors including complete-matched, bottom-eliminated, and top-reserved.
%We use \texttt{HealthyGo} as a case study to illustrate the optimal threshold matching rules and corroborate the analytical results.
\end{abstract}

\IEEEpeerreviewmaketitle
%%%%%%%%%%%%%%%%%%%%%%%%
%%%%%%%%%%%%%%%%%%%%%%%%

\section{Introduction}

%As the number of cyber and physical devices are connected to the existing Internet infrastructure, the Internet of Things (IoT) provides connectivity and network solutions to emerging applications such as smart homes, enterprises, and cities \cite{perera2015emerging}.
%

IoT system generates tons of data that has a significant amount of valuable knowledge hidden.
Data collected for fulfilling the individual tasks are often either deleted or stored and locked down in independent data silos.
On the one hand, the data owners generally lack knowledge and techniques to conduct knowledge discovery from their data and can only gain access to their own data which has little value with respect to knowledge discovery \cite{perera2017sensing}.
On the other hand, the data consumers have abilities to discover the hidden knowledge accurately and efficiently, but they may not have a sufficient amount of targeted data or the rights to access to the data due to privacy concerns.
This asymmetric relationship between the data owners and the data consumers leads to a potential market opportunity for the IoT data sharing.

In this work, we focus on massive IoT networks with a sufficiently large number of data owners and data consumers and develop a two-sided matching market model with a monopolist trading platform for the IoT data sharing.
This model captures the different roles of data owners and data consumers in dealing with data and conceptualizes the data owners as the sellers and the data consumers as the buyers.
Sellers possess IoT data generated during the primary usages of different IoT products to sell and have different preferences for the payoffs from participation.
On the other hand, buyers offer rewards to the sellers for sharing their data and have different requirements on the data, including, for example, its features (e.g., category, history, quality, and quantity of the data), the users' engagements with the IoT system (e.g., the usage frequency of IoT products), and the users' personal information (e.g., life styles and financial status).
We consider that there is a monopolist intermediary matching platform (platform) that takes advantage of this market opportunity to offer a matching mechanism that enables and encourages interactions between two sides of participants and matches sellers and buyers based on their private information that summarizes their preferences.
%
% The monopolist platform (platform) plays a role to enable and encourage interactions between two sides of participants and matches the owners and the buyers based on their information and preferences by charging both sides with an appropriate price for offering a matching mechanism.
%
%The charged price can then generate revenue for the platform.
%

The matching mechanism consists of a matching rule and a payment rule.
The matching rule matches the owners and the buyers by specifying the datasets that can provide the highest utility for the buyers and determines the rewards the owners can receive from the buyers.
As a result, the matching rules produce welfare for owners and buyers.
The payment rule specifies the payments from each side of the market once a match is established that generate revenues to the platform.
The proposed model considers the interactions between a large number of owners and buyers and study the matching mechanisms for revenue and welfare maximization \cite{hagiu2014information}, respectively.
We describe a class of threshold matching rules and show that under certain assumptions both the welfare-maximizing and the revenue-maximizing matching mechanisms have optimal cut-off matching rules.
We study different matching behaviors and characterize different matching patterns based on the joint marginal effect of matching two individual participants on the platform's mechanism design goal. 
An numerical example is provided to illustrate the theoretical analysis of this work.
%
%
% The mechanism generates a revenue to the platform by charging both sides with an appropriate price.
%
% The matching rule determines the datasets that can provide the highest utility for the buyers.
% Since the matching is reciprocal, the data shared according to the matching rule, on the other hand, determines the rewards the owners can receive from the buyers.
% %
% In other words, the matching rules can generate welfare over the IoT network.

%\cite{zhang2019optimal}

This work is related to matching mechanism design for matching two-sided market (e.g., one-to-one matching \cite{johnson2013matching} and many-to-many matching \cite{gomes2016many}).
There is a rich literature on the economic analysis and pricing schemes of the market model for data collection in IoT networks \cite{aazam2015fog, perera2014sensing, al2013priced} based on a variety of approaches, including smart data pricing scheme \cite{sen2015smart} such as sealed-bit auctions \cite{sun2006wireless}.
Also, utility maximization-based pricing schemes have also been studied \cite{lee2005non, lee2006jointly}.
For example, in \cite{farooq2018optimal}, an optimal dynamic spectrum reservation contract has been designed for mission-critical IoT systems, in which an advance payment is made at the time of reservation and a rebate is made if the reservation is released. The IoT applications are incentivized to reveal the true application type and it has been shown that the incentive compatible mechanism leads to an efficient utilization of the spectrum as well as a greater revenueability of the IoT network operator. 
\section{IoT Data Market}\label{Sec_2}

\subsection{Market Model}

\subsubsection{Information structure}

There are two sides in the IoT data market: the \textit{seller} side ($S$) consisting of participants that are data owners, and the \textit{buyer} side ($B$) consisting of participants who are the data buyers.
We use $K$ to denote one side $K\in \{S,B\}$ and $\bar{K}$ to denote its opposite $\bar{K}\in \{S,B\}\backslash\{K\}$.
%
% $$AAAAAAA$$
% In the scenario of IoT, the data market has two sides: the seller side ($S$) constituting of (market) participants that are the IoT data owners,  and the buyer side ($B$) constituting of participants who are the IoT data buyers.
%
%
We consider a massive IoT networks, in which there is a sufficiently large number of participants on both sides of the market, and each participant has no market power. As a result, we model the population of each side of the market as a unit-mass continuum of participants over $[0,1]$.
%
% the population of each side of the market can be modeled by a unit-mass continuum of participants indexed by $i\in[0,1]$.
%
Each participant from side $K$ has a \textit{type} $\lambda^{K}\in \Lambda^{K} = [\underline{\lambda}^{K}, \bar{\lambda}^{K}]\in \mathbb{R}$.
We assume that $\lambda^{K}$ is drawn independently from a continuous distribution $F^{K}(\lambda^{K})$ with density $f^{K}(\lambda^{K})$.
Each seller's type $\lambda^{S}\in \lambda^{S}$ summarizes the feature of her data (including, e.g., quality, categories, or cost performance). 
%and her expectation of rewards from the buyer, or the cost performance.
%
On the other hand, each buyer's type $\lambda^{B}\in \lambda^{B}$ summarizes her requirements of data (e.g., quality, quantity) and the offers she can provide to the sellers.
The type of each participant from one side also reflects the attractiveness of the participant seen by those in the other side of the market.
We assume that for any two types $\lambda^{K}_{h}\in \lambda^{K}$ and $\lambda^{K}_{\ell}\in \lambda^{K}$, participants of $\lambda^{K}_{h}$ is more attractive than those of $\lambda^{K}_{\ell}$ if $\lambda^{K}_{\ell}<\lambda^{K}_{h}$, e.g.,
buyers prefer to be matched to sellers of higher cost performance and sellers prefer to be matched to buyers of better offers.
Informally, let $C(\lambda^{S}, \lambda^{B})$ represent the \textit{contribution} generated by matching \textit{one} seller of $\lambda^{S}$ to \textit{one} buyer of $\lambda^{B}$ to the platform's goal.
The following assumption specifies a monotone property of $C(\lambda^{S}, \lambda^{B})$.
\begin{assumption}\label{assp:order_of_preference}
Let $\lambda^{K}_{h}$, $\lambda^{K}_{\ell}\in \Lambda^{K}$ be any two types from side $K\in\{S,B\}$, with $\lambda^{K}_{\ell}\leq \lambda^{K}_{h}$. Then, for $a\in \{h,\ell\}$ and $\bar{a}\in \{h,\ell\}\backslash\{a\}$,
$$
C(\lambda^{S}_{\ell}, \lambda^{B}_{\ell})\leq C(\lambda^{S}_{a}, \lambda^{B}_{\bar{a}}) \leq C(\lambda^{S}_{h}, \lambda^{B}_{h}).
$$
\end{assumption}

Assumption \ref{assp:order_of_preference} claims that the matching mechanism satisfies a monotone property, i.e., matching a seller of higher cost performance to a buyer of better offer generates higher contribution than matching participants of lower cost performance or lower offer.
%
%Following Assumption \ref{assp:order_of_preference}, we consider that all the participants are rational in the sense that a participant of higher $\lambda^{K}$ is more willing to participate in the data market than a participant of lower $\lambda^{K}_{\ell}$.
%
%In other words, participants of higher type has higher expectations from trading the data.

However, each participant's type is her own private information and the matching mechanism requires each participant to report her type to the platform through a \textit{message}.
Due to this information asymmetry, each participant can game the system to her own advantage by strategically reporting her type to the platform.
Let $\Omega^{K}$ denote the set of messages a participant from $K$ can choose.
Define participant's \textit{reporting strategy} $\alpha^{K}: \Lambda^{K}\mapsto \Omega^{K}$, such that a participant reports a message $m^{K} = \alpha^{K}(\lambda^{K})$ when her true type is $\lambda^{K}$.
We consider the direct mechanism, in which participants directly reveal their type (truthfully or falsified), i.e, $\Omega^{K} = \Lambda^{K}$.
We say the participant's reporting strategy is \textit{truthful} if $\alpha^{K}(\lambda^{K})=\lambda^{K}$, for all $\lambda^{K}\in\Lambda^{K}$, $K\in\{S,B\}$.
%

%
% to pay to the platform for being matching to $\lambda^{\bar{K}}_{h}$ than $\lambda^{\bar{K}}_{l}$, for any $\lambda^{\bar{K}}_{l}\leq \lambda^{\bar{K}}_{h}$.
% %
% Additionally, we assume that a participant of higher $\lambda^{K}_{h}$ is more willing to 

% values the matching more than a participant of lower $\lambda^{K}_{\ell}$ does when they are matched to the same participant of side $K$ of relatively high 
%

\subsubsection{Mechanism rules}

The data market is driven by the platform (he), who operates as a data market maker that provides a matching mechanism consisting of a matching rule that matches participants from one side to those from the other side and a payment rule that charges participants from each matched participant by a proper price.
Let $\sigma^{K}: \Lambda^{K}\mapsto P(\Lambda^{\bar{K}})$ denote the \textit{matching rule} for side $K\in\{S,B\}$, where $P(\Lambda^{\bar{K}})$ denotes the power set of $\Lambda^{\bar{K}}$, such that $\sigma^{K}(\hat{\lambda}^{K})$ gives the set of types on the side $\bar{K}$ that a participant of reported type $\hat{\lambda}^{K}\in \Lambda^{K}$ is matched to.
Let $\Xi^{K}\subseteq \Lambda^{K}$ denote any subset of $\Lambda^{K}$.
This point-to-set mapping captures the fact that a database may be sold to multiple buyers of different types and a buyer may buy multiple databases from sellers of different types.
Let $\phi^{K}: \Lambda^{K} \mapsto \mathbb{R}_{+}$ denote the \textit{payment rule}, such that $\phi^{K}(\hat{\lambda}^{K})$ specifies a payment that the participant reporting $\hat{\lambda}^{K}$ needs to pay to the platform for being matched.
Let $\bm{\sigma}=\{\sigma^{S}, \sigma^{B}\}$ and $\bm{\phi} = \{\phi^{S}, \phi^{B}\}$.
The matching rule $\bm{\sigma}$ is \textit{feasible} if it satisfies a reciprocal relationship: if a participant of $\lambda^{K}$ is matched to a participant of $\lambda^{\bar{K}}$ from the opposite side, then the participant of $\lambda^{\bar{K}}$ must also be matched to the participant of $\lambda^{K}$, i.e., for $K\in\{S,B\}$, $\bar{K}\in \{S,B\}\backslash\{K\}$,
\begin{equation}\label{eq:reciprocal}
     \lambda^{K}\in\sigma^{\bar{K}}(\lambda^{\bar{K}}), \textit{ if and only if, } \lambda^{\bar{K}}\in \sigma^{K}(\lambda^{K}). 
\end{equation}
The reciprocal relationship (\ref{eq:reciprocal}) holds when $\lambda^{K}$ and $\lambda^{\bar{K}}$ are the reports sent by the matched participants.

\subsubsection{Utility and Payoff} 

In the data market, the sellers are motivated to participate by the reward they may obtain by sharing their data to the buyers, while the buyers are motivated by the benefits of utilizing the data.
Basically, the buyers acknowledge the (potential) value of data and recognize the importance of providing the reward to motivate the sellers to share their data.
%
% Let $R^{S}(\cdot | \sigma^{S}): P(\Lambda^{\bar{S}})\mapsto \mathbb{S}_{+}$ denote the (expected) \textit{reward} function, such that a seller of $\lambda^{S}$ is offered by the reward $R^{S}(\Xi^{B}|\sigma^{S})$ when she is matched to $\Xi^{B}$ by $\sigma^{S}$.
%
%
Let $\gamma^{B}: \Lambda^{B} \mapsto \Gamma \subset \mathbb{R}_{+}$ denote the reward, such that $r=\gamma^{B}(\lambda^{B})$ is the reward offered by the buyer of type $\lambda^{B}$ to each of her matched sellers.
%
% We assume that the reward function $\gamma$ is fixed and all the participants and the platform agree on it, i.e., if a buyer reports $\hat{\lambda}^{B}$, then she needs to provide the reward $\gamma(\hat{\lambda}^{B})$.
%
Let $M^{K}:\mathbb{R}_{+} \times \Lambda^{K}\mapsto \mathbb{R}_{+}$ denote the monetary evaluation of the reward by participants from $K$, $K\in\{S,B\}$.
Hence, $M^{S}(r, \lambda^{S})$ is the monetary value of the reward $r$ perceived by the seller of $\lambda^{S}$. 
% We assume that $M^{K}$ and $\gamma^{B}$ are continuous and differentiable and $M^{B}(\gamma^{B}(\lambda^{B}, \lambda^{S}) \leq M^{S}(\gamma^{B}(\lambda^{B}),\lambda^{S})$, i.e., the cost of offering $r=\gamma^{B}(\lambda^{B})$ by type-$\lambda^{B}$ (might be falsified) buyer is less than the monetary value of $r$ as seen as a reward by the type-$\lambda^{S}$ seller, 
% This assumption is based on the setting that the reward may not be a direct monetary transfer but through the terms of services, products, or vouchers from the buyers' business, in which the buyers' cost is lower than the tagged price.
%
On the other hand, the shared data produces value to the buyers from utilizing the data.
Define the \textit{data value} as $\gamma^{S}:\Lambda^{S}\mapsto \mathbb{R}_{+}$, such that $M^{B}(\gamma^{S}(\lambda^{S}), \lambda^{B})$ is the data value provided by seller of $\lambda^{S}$ that is recognized by the buyer of $\lambda^{B}$.
%
% $v: \Lambda^{S}\times \Lambda^{B} \mapsto \mathbb{R}_{+}$, such that $v(\lambda^{S}, \lambda^{B})$ is the data value perceived by the buyer of type $\lambda^{B}$ when the data is from the seller of $\lambda^{S}$ (might be falsified).
% a buyer of $\lambda^{B}$ can generate $v(\lambda^{S}, \lambda^{B})$ from utilizing the data from type-$\lambda^{S}$ seller.
%
% Because $v(\lambda^{S}, \lambda^{B})$ represents the true value of utilizing the data, it depends on the true types of the participants and is independent of the reports.
% the reward received by the seller of $\lambda^{S}$, who reports $\hat{\lambda}^{S}$, from the matched buyer of reported type $\hat{\lambda}^{B}$.
%
We assume that $M^{K}$ and $\gamma^{K}$ are continuous and differentiable and they are fixed and known by all participants and the platform.
%
% We assume that the reward function $\gamma^{B}$ and the data value function $\gamma^{S}$ are fixed and known by all the participants and the platform.
Additionally, we assume that $\gamma^{B}(\lambda^{B})\leq M^{S}(\gamma^{B}(\lambda^{B}),\lambda^{S})$, i.e., the cost of offering $r=\gamma^{B}(\lambda^{B})$ by type-$\lambda^{B}$ buyer is less than the monetary value of $r$ as seen as a reward by the type-$\lambda^{S}$ seller.
This assumption is based on the setting that the reward may not be a direct monetary transfer but through the terms of services, products, or vouchers from the buyers' business, in which the buyers' cost is in general lower than the tagged price.
%
% all the participants and the platform agree on it, i.e., if a buyer reports $\hat{\lambda}^{B}$, then she needs to provide the reward $\gamma(\hat{\lambda}^{B})$.

Given $\sigma^{K}$, define the \textit{utility} function as $u^{K}(\cdot, \cdot| \sigma^{K}, \alpha^{K}): \Lambda^{K}\times P(\Lambda^{\bar{K}})\mapsto \mathbb{R}_{+}$, such that $u^{K}(\lambda^{K}, \Xi^{\bar{K}}|\sigma^{K}, \alpha^{K})$ specifies the utility for a participant from $K$ when her true type is $\lambda^{K}$ and she adopts reporting strategy $\alpha^{K}$.
Let $R^{S}(\lambda^{S}, \lambda^{B}|\sigma^{S}, \alpha^{S}) = M^{S}(\gamma^{B}(\lambda^{B}), \lambda^{S})$ and $R^{B}(\lambda^{B}, \lambda^{S}|\sigma^{B}, \alpha^{B}) = M^{B}(\gamma^{S}(\lambda^{S}), \lambda^{B}) - \gamma^{B}(\alpha^{B}(\lambda^{B}))$.
Then, for $K\in \{S,B\}$, $\bar{K}\in\{S,B\}\backslash{\{K\}}$,

\begin{equation}\label{eq:utility}
    u^{K}(\lambda^{K}, \Xi^{\bar{K}}|\sigma^{K}, \alpha^{K}) = \int_{\Xi^{\bar{K}}} R^{K}(\lambda^{K},x|\sigma^{K},\alpha^{K}) f^{\bar{K}}(x) dx.
\end{equation}
%
% \begin{equation}\label{eq:utility_buyer}
%     u^{B}(\lambda^{B}, \Xi^{S}|\sigma^{S}) = \int_{\Xi^{S}}\Big[v(x,\lambda^{B}) - M^{B}(\gamma(x,\lambda^{B});\lambda^{B})  \Big]f^{S}(x)d x.
% \end{equation}
%
We have the following assumption of $u^{K}$.

\begin{assumption}\label{assp:utility}
The utility $u^{K}(\lambda^{K}, \sigma^{K}(\hat{\lambda}^{K})|\sigma^{K}, \alpha^{K})$ is a continuous in $\lambda^{K}$ and non-decreasing in report $\hat{\lambda}^{K}$.
%
% The utility $u^{K}$ is a continuous function of $\lambda^{K}$ and is non-decreasing in $\lambda^{K}$, with $u^{K}(\lambda^{K}, \emptyset|\sigma^{S})=0$, for $K\in\{S,B\}$.
\end{assumption}

It is straightforward to verify that the utility $u^{K}$ satisfying Assumption \ref{assp:utility} coincides with the monotonicity of $C$ in Assumption \ref{assp:order_of_preference}.

Define \textit{payoff function}, $J^{K}(\cdot| \sigma^{K}, \phi^{K}, \alpha^{K}_{i}): \Lambda^{K}\times \mathbb{R}_{+} \times P(\Lambda^{\bar{K}})\mapsto \mathbb{R}$, of a participant of $\lambda^{K}$ from $K$, when she adopts reporting strategy $\alpha^{K}$, 
\begin{equation}\label{eq:payoff_function}
    \begin{split}
        J^{K}(\lambda^{K}, p^{K}, \Xi^{\bar{K}}\big|&\sigma^{K}, \phi^{K}, \alpha^{K}_{i}) = u^{K}\Big(\lambda^{K}, \Xi^{\bar{K}}\big|\sigma^{S}\Big)-p^{K}.
    \end{split}
\end{equation}

To simplify the notations, we remove the rules and strategy $\sigma^{K}$, $\phi^{K}$, and $\alpha^{K}$ in the left-hand side of $J^{K}$, unless otherwise stated.

%%%%%%%%%%%%%%%%%%%%%%%%%%%%%%%
\section{Matching Mechanism}

In the letter, we focus on the \textit{anonymous} mechanism design.
Specifically, the set of participants from side $\bar{K}\in \{S,B\}$ is matched to a participant of $\lambda^{K}$ from $K\in\{S,B\}\backslash\{\bar{K}\}$ and the associated payments depend only on the reported type $\hat{\lambda}^{K} = \sigma^{K}(\lambda^{K})$.
Let $\sigma^{K}_{i}(\lambda^{K}_{i})\in \Lambda^{K}$ and $\sigma^{K}_{j}(\lambda^{K}_{j})\in \Lambda^{K}$ be the types of two participants $i$ and $j$, respectively.
If $\sigma^{K}_{i}(\lambda^{K}_{i}) = \sigma^{K}_{j}(\lambda^{K}_{j})$,then both participants $i$ and $j$ receive the same matched set $\Xi^{\bar{K}}$ and payment $p^{K}$.
This is because the matching rule $\sigma^{K}$ and payment rule $\phi^{K}$, respectively, specify a match and a payment that depend only on her reported type and are independent of other participants from the same side $K$.
As a result, the identity of each participant can be fully characterized by her reported type.
%
%Hence, we omit the index $i$ in the reporting strategy, unless otherwise stated.

Due to the revelation principle \cite{myerson1981optimal}, our model considers the direct mechanism and requires the matching mechanism to incentivize all the participants from both sides to report their types truthfully.
This is established by imposing the \textit{incentive compatibility} (IC) constraint to the platform's mechanism design problem.

\begin{definition}[\textbf{Individual Compatibility}]\label{def:incentive_compatibility}
The matching mechanism $<\bm{\sigma}, \bm{\phi}>$ is \text{incentive compatible} if 
\begin{equation}\label{eq:def_IC}
    J^{K}\Big(\lambda^{K}, \phi^{K}\big( \lambda^{K}\big), \sigma^{K}\big( \lambda^{K} \big) \Big)\geq J^{K}\Big(\lambda^{K}, \phi^{K}\big( \hat{\lambda}^{K}\big), \sigma^{K}\big( \hat{\lambda}^{K} \big) \Big),
\end{equation}
for any $\lambda^{K}$, $\hat{\lambda}^{K}\in \Lambda^{K}$.
\end{definition}

In incentive compatible matching mechanism, the truthful reporting strategy is the (weakly) dominant strategy for each participant.
The following definition shows a first-order condition for incentive compatibility (ICFOC) based on envelope theorem.

\begin{definition}[\textbf{ICFOC}]\label{def:ICFOC}
A matching mechanism $<\bm{\sigma}, \bm{\phi}>$ satisfies ICFOC if $J^{K}\Big(\lambda^{K}, \phi^{K}\big( \lambda^{K}\big), \sigma^{K}\big( \lambda^{K} \big) \Big)$ is a continuous and differentiable function of $\lambda^{K}$ with the derivative given almost everywhere by 
\begin{equation}\label{eq:def_ICFOC}
    \begin{split}
        \frac{\partial J^{K}\big( x, \phi^{K}(x), \sigma^{K}(x)\big)}{\partial x }\Big|_{x= \lambda^{K}} = \frac{\partial}{\partial x}u^{K}(x, \sigma^{K}(\lambda^{K})|\sigma^{K})\Big|_{x=\lambda^{K}}.
    \end{split}
\end{equation}

%
%\textbf{\textcolor{blue}{If the number of data matters for the buyer, what will happen to the incentive compatibility?}}
%

%
\end{definition}

The ICFOC condition in Definition \ref{def:ICFOC} implies that the payoff of each matched participant can be characterized by the matching rule $\sigma^{K}$, $K\in\{S,B\}$.
Define
\begin{equation}\label{eq:FOC_term}
D^{K}(\lambda^{K}) =    \frac{\partial}{\partial x}u^{K}(x, \sigma^{K}(\lambda^{K})|\sigma^{K})\Big|_{x=\lambda^{K}},
\end{equation}
%
% \begin{lemma}
% Suppose Assumption \ref{assp:utility} holds. If  Then, $D^{K}$ is non-decreasing in $\lambda^{K}$.
% \end{lemma}
%
and
\begin{equation}\label{eq:information_rent}
    Q^{K}(\lambda^{K}) = \int^{\lambda^{K}}_{\tilde{\lambda}^{K}} D^{K}(x) dx,
\end{equation}
where $\tilde{\lambda}^{K}\in \Lambda^{K}$ is some arbitrarily fixed type. 
Here, $Q^{K}(\lambda^{K})$ can be interpreted as the advantage (an economic rent) a participant of $\lambda^{K}$ has over some type $\tilde{\lambda}^{K}$ due to the platform's not knowing that her true type is $\lambda^{K}$.
For example, if $J^{K}$ is an increasing function of $\lambda^{K}$, then $Q^{K}(\lambda^{K})$ with $\tilde{\lambda}^{K} = \underline{\lambda}^{K}$ is the maximum advantage a participant of $\lambda^{K}$ holds. 
In this example, it is straightforward to see that participants of $\bar{\lambda}^{K}$ have no incentive to misreport their true type, while participants of $\underline{\lambda}^{K}$ have incentive to misreport $\underline{\lambda}^{K}$ as $\bar{\lambda}^{K}$ because she has no advantage by truthful reporting compared to any type $\lambda^{K}> \underline{\lambda}^{K}$.
Hence, due to the ICFOC, guaranteeing the IC constraint requires to craft the payment rule such that advantages (or disadvantages) that different types hold in the utilities can be balanced.
The following proposition shows the necessary and sufficient conditions for matching mechanisms being incentive compatible.

\begin{proposition}\label{prop:IC}
Suppose Assumption \ref{assp:utility} holds.
A matching mechanism $<\bm{\sigma}, \bm{\phi}>$ is incentive compatible if and only if the payment rule is constructed as follows, for all $K\in\{S,B\}$,
\begin{equation}\label{eq:construct_payment}
        \begin{aligned}
        \phi^{K}(\lambda^{K}) =& u^{K}(\lambda^{K},\sigma^{K}(\lambda^{K})) - \int^{\lambda^{K}}_{\underline{\lambda}^{K}}D^{K}(x) dx \\
        -& J^{K}(\underline{\lambda}^{K}, \phi^{K}(\underline{\lambda}^{K}), \sigma^{K}(\underline{\lambda}^{K})),
        \end{aligned}
    \end{equation}
    where $D^{K}$ is given in (\ref{eq:FOC_term}).
%
% \begin{itemize}
%     \item[(i)] the payment rule is constructed as follows:
    
%     % $$
%     % J^{K}(\underline{\lambda}^{K}, \phi^{K}(\underline{\lambda}^{K}), \sigma^{K}(\underline{\lambda}^{K}))\geq 0;
%     % $$
%     %
%     % \item[(ii)] the matching rule $\sigma^{K}$ is nondecreasing in $\lambda^{K}$.
%     %
% \end{itemize}
%
\end{proposition}

The construction of $\phi^{K}$ in (\ref{eq:construct_payment}) can be interpreted as the utility of truthfully reporting $\lambda^{K}$, advantage of type $\lambda^{K}$ over $\underline{\lambda}^{K}$, and the payoff of $\underline{\lambda}^{K}$.
Proposition \ref{prop:IC} also requires that the payoff of the lowest type $\underline{\lambda}^{K}$ is designed by choosing a constant.
Given the IC constraint, we rewrite the payoff in (\ref{eq:payoff_function}) as $J^{K}(\lambda^{K}|\sigma^{K}, \phi^{K})$.

Another important constraint in our matching mechanism design is the \textit{individual rationality} (IR) constraint.
In addition to the IC constraint, the IR constraint requires that the payoff of each participant of $\lambda^{K}$ is non-negative, i.e., $J^{K}(\lambda^{K}|\sigma^{K}, \phi^{K})\geq 0$, for all $\lambda^{K}\in \Lambda^{K}$, $K\in\{S,B\}$.

% \begin{corollary}\label{corollary:IR}
% %
% Suppose Assumption \ref{assp:utility} holds.
% The incentive compatible matching mechanism $<\bm{\sigma}, \bm{\phi}>$ is individually rational if and only if, for $K\in\{S,B\}$
% %
% \begin{equation}\label{eq:corollary_IR}
%     J^{K}(\underline{\lambda}^{K}|\sigma^{K}, \phi^{K})\geq 0.
% \end{equation}
% %
% \end{corollary}

%%%%%%%%%
\subsection{Revenue and Welfare Maximization}

The expected social welfare generated by the matching mechanism is given as
\begin{equation}\label{eq:revenue}
    Z_{W}(\bm{\sigma}, \bm{\phi}) = \sum_{K\in\{S,B\}} \int_{\Lambda^{K}} u^{K}(x, \sigma^{K}(x)|\sigma^{K})f^{K}(x) dx,
\end{equation}
and the expected revenue obtained by the platform is given as
\begin{equation}\label{eq:social_welfare}
    Z_{R}(\bm{\sigma}, \bm{\phi}) = \sum_{K\in\{S,B\}} \int_{\Lambda^{K}} \phi^{K}(x) f^{K}(x) dx.
\end{equation}

By substituting the construction of $\phi^{K}$ in (\ref{eq:construct_payment}) into (\ref{eq:social_welfare}), we have
\begin{equation}\label{eq:revenue_v1}
    \begin{aligned}
    Z_{R}(\bm{\sigma}, \bm{\phi}) =& \sum_{K\in\{S,B\}} \int_{\Lambda^{K}}\Big[u^{K}(x, \sigma^{K}(x)|\sigma^{K}) -  \int^{x}_{\underline{\lambda}^{K}} D^{K}(r) dr   \\
    &-J^{K}(\underline{\lambda}^{K}|\sigma^{K}, \phi^{K})\Big]f^{K}(x) dx.
    \end{aligned}
\end{equation}

\begin{corollary}\label{corollary:revenue_lower}
Suppose Assumption \ref{assp:utility} holds.
In the incentive compatible matching mechanism $<\bm{\sigma}, \bm{\phi}>$, the followings hold.
\begin{itemize}
    \item[(i)] The mechanism is individually rational if and only if, for $K\in\{S,B\}$,
\begin{equation}\label{eq:corollary_IR}
    J^{K}(\underline{\lambda}^{K}|\sigma^{K}, \phi^{K})\geq 0.
\end{equation}
\item[(ii)] In the matching mechanism that maximizes the social welfare $Z_{W}(\bm{\sigma},\bm{\phi})$, $J^{K}(\underline{\lambda}^{K}|\sigma^{K}, \phi^{K})=0$.
\end{itemize}
\end{corollary}

%%%%%%%%%%%%%%%%%%%%%%%%%%%%%%%%%%%%%%
%\section{Optimal Matching Mechanism}
\subsection{Cut-off Rule}

In this section, we describe a class of matching mechanism based on a cut-off rule.
Let $\tau^{K}(\cdot| \delta^{K}): \Lambda^{K}\mapsto \Lambda^{\bar{K}}$, for $K\in \{S,B\}$, be the \textit{cut-off} function with the \textit{threshold} $\delta^{K} \in \Lambda^{K}$, such that the matching rule is defined as follows: 
\begin{equation}\label{eq:cut_off}
    \begin{aligned}
    \sigma^{K}(\lambda^{K})=& [\tau^{K}(\lambda^{K}|\delta^{K}),\;\; \bar{\lambda}^{K}]\mathbf{1}_{\{\lambda^{K}\geq \delta^{K}\}} + \emptyset \mathbf{1}_{\{\lambda^{K}< \delta^{K}\}},
    \end{aligned}
\end{equation}
where $\mathbf{1}_{\{\cdot\}}$ is the indicator function.
The matching rule in (\ref{eq:cut_off}) can be interpreted as, for example, the seller with overall data quality characterized by $\lambda^{S}$ that is at least $\delta^{S}$ is matched to buyers whose data requirements (and the associated reward) characterized by $\lambda^{B}$ is at least $\tau^{S}(\lambda^{S}|\delta^{S})$; otherwise the seller is not matched. Similar interpretation can be made for the buyers' side.
%

% \textbf{\textcolor{red}{Double check the following tradeoffs.}}
% We consider a tradeoff between data seller and data buyer: sellers of higher type (e.g., higher quality data or cost performance) may prefer to be matched to buyers of a wider range of types (from lower to higher requirements of data) to receive more rewards, while the buyers of higher type (e.g., higher requirements of data cost performance) may prefer to be matched to sellers of relatively higher types to extract more value from the data.
% %
% This tradeoff, however, may violate the fundamental reciprocal relationship of $\sigma^{K}$ in (\ref{eq:reciprocal}).
%
The following definition shows the conditions of the cut-off based matching rules such that the reciprocal relationship in (\ref{eq:reciprocal}) is satisfied.

\begin{definition}[\textbf{Cut-Off Matching Rule}]\label{def:cut_off}
The cutoff rule in (\ref{eq:cut_off}) is a feasible matching rule if (i) $\tau^{K}(\lambda^{K}|\delta^{K})=\inf\{\lambda^{\bar{K}}: \tau^{\bar{K}}(\lambda^{\bar{K}}|\delta^{\bar{K}})\leq \lambda^{K} \}$ ; (ii) $\delta^{K}=\inf\{\lambda^{K}:\tau^{\bar{K}}(\lambda^{\bar{K}}|\delta^{\bar{K                               }})\leq \lambda^{K}, \lambda^{\bar{K}}\in \Lambda^{\bar{K}}\}$; (iii) $\tau^{K}(\lambda^{K}|\delta^{K})$ is non-increasing in $\lambda^{K}$. 
\end{definition}

% The following lemma shows the conditions such that the cut-off based matching rules satisfy the reciprocal relationship.

% \begin{lemma}\label{lemma:reciprocal_cut_off}
% %
% The matching rule $\sigma^{K}$, for $K\in\{S,B\}$, is a cut-off rule with $\tau^{K}(\cdot|\sigma^{K})$ is reciprocal if and only if : \textcolor{red}{(i) $R^{B}(\lambda^{B}, \lambda^{S})$ is non-decreasing in $\lambda^{B}$ with $R^{B}(\lambda^{B}, \tilde{\lambda}^{S})=0$ and $R^{B}(\lambda^{B}, \bar{\lambda}^{S})>0$, for some $\tilde{\lambda}^{B}\geq \tau^{B}(\lambda^{B}|\delta^{B})$ \textcolor{red}{($\tilde{\lambda}^{B}\geq \bar{\lambda}^{K}$)}, all $\bar{\lambda}^{S}> \tilde{\lambda}^{B}$;}
% %$M^{B}(\gamma(\lambda^{S}, \lambda^{B}); \lambda^{B})$ is strictly increasing in $\lambda^{B}$ with $M^{B}(\gamma(\lambda^{S}, \tilde{\lambda}^{B}); \lambda^{B})=0$, for some $\tilde{\lambda}^{B}> \tau^{B}(\lambda^{B}|\delta^{B})$; 
% %
% (ii) $\delta^{K}=\min\{\lambda^{K}:\tau^{\bar{K}}(\lambda^{\bar{K}})\leq \lambda^{K}, \lambda^{\bar{K}}\in \Lambda^{\bar{K}}\}$; (iii) $\tau^{K}(\lambda^{K}|\delta^{K})$ is non-increasing in $\lambda^{K}$.  
% %
% \end{lemma}

In Definition \ref{def:cut_off}, the conditions (i) and (ii) specify the reciprocal relationships of the cut-off functions and the thresholds between sides $S$ and $B$ and the condition (iii) specifies the monotonicity of the cut-off function, such that the matching rule defined in (\ref{eq:cut_off}) is feasible, i.e. (\ref{eq:reciprocal}) is satisfied.

% Condition (i) requires that the buyers of $\lambda^{B}$ receives no reward from the matched sellers of sufficiently small $\tilde{\lambda}^{S}\geq\tau^{B}(\lambda^{B}|\delta^{B})$, i.e., $v(\tilde{\lambda}^{S}, \lambda^{B}) = M^{B}(\gamma(\tilde{\lambda}^{S}, \lambda^{B}))$.
% %
% This guarantees that matching to lower-type sellers (e.g., with low-quality data) impose zero loss to the buyers.
% %
% Condition (ii) is from the inverse of the matching rule and the reciprocal relationship.
%

\begin{lemma}\label{lemma:existence_cut_off}
Suppose Assumption \ref{assp:utility} holds. Then, matching rules of the welfare-maximizing and the revenue-maximizing mechanisms are cut-off rules.
\end{lemma}

Hence, the platform's mechanism design problem is to maximize the social welfare or his revenue by determining the cut-off functions $\bm{\tau}=\{\tau^{S}, \tau^{B}\}$ with thresholds $\bm{\delta}=\{\delta^{S}, \delta^{B}\}$ and the payment rule $\bm{\phi}$, i.e., for $Y\in\{W,R\}$,
\begin{equation}\label{eq:platform_problem}
    \begin{split}
        &\max_{\bm{\tau},\bm{\delta}, \bm{\phi}}\;\; Z_{Y}(\{\bm{\tau}, \bm{\delta}\}, \bm{\phi})\\
        \text{s.t. }&\text{ IC, IR, and (i)-(iii) in  Definition \ref{def:cut_off}}.
    \end{split}
\end{equation}

% %%%%%%%%%
% \begin{figure}
%   \centering
%     \includegraphics[width=0.4\textwidth]{Figure/Platform_match}
%     \caption{Data trading model: Each data owner and each data buyer report their data categories (true values) and types (reported values) to the platform. At the first matching stage, the platform matches the groups of owners and the groups of buyers according to their data category. At the second matching stage, the platform uses the reported types and returns the optimal matching rules to the market participants.}
%     \label{Fig_platform}
% \end{figure}

% \textcolor{orange}{\textbf{Figure} \ref{Fig_platform}.}

%%%%%%%%%

%%%%%%%%%%%%%%%%%%%%%%%%%%%%%%%%%%%%%%
\section{Characterization the Matching Mechanism}

From Assumption \ref{assp:utility}, it is straightforward to see that both sellers and buyers prefer a match to opponents of higher types, i.e., $\tau^{K}(\lambda^{K}|\delta^{K})\leq \bar{\lambda}^{\bar{K}}$, for all $\lambda^{K}\in \Lambda^{K}$, $K\in \{S,B\}$, and $\bar{K}\in \{S,B\}\backslash\{K\}$, given that IC and IR constraints are satisfied.
Their preferences over the opponents of lower types, however, may exhibit inconsistency between two sides.
Suppose a seller of $\lambda^{S}\in \Lambda^{S}$ is matched to a buyer of $\lambda^{B}$.
Then the seller receives a reward $M^{S}(\gamma^{B}(\gamma^{B}), \lambda^{S})$ and the buyer receives a reward $M^{B}(\gamma^{S}(\lambda^{S}), \lambda^{B})-\gamma^{B}(\lambda^{B})$.
%
% Suppose a seller of $\lambda^{S}\in \Lambda^{S}$ is matched to two buyers $i$ and $j$, of $\lambda^{B}_{i}\in \Lambda^{B}$ and $\lambda^{B}_{j}\in \Lambda^{B}$, respectively.
% %
% Then buyers $i$ and $j$ receive data value $M^{B}(\gamma^{S}(\lambda^{S}), \lambda^{B}_{i})$ and $M^{B}(\gamma^{S}(\lambda^{S}), \lambda^{B}_{j})$, respectively, and the seller receives rewards $\sum_{k\in\{i,j\}}M^{S}(\gamma^{B}(\lambda^{B}_{k}), \lambda^{S})-M^{S}(\gamma^{B}(\lambda^{B}_{k}), \lambda^{S}_{k})$.
%
Since $\gamma^{B}(\lambda^{B})$ is non-negative and the same data can be copied multiple times without additional cost, the seller also prefers to be matched to buyers of a wide range of types.
The buyers, on the other hand, may receives negative reward by being matched to lower types.
As a result, buyers of some types may not prefer to be matched to sellers of (sufficiently) low types.
We define the following matching patterns when the platform adopts the cut-off matching rule defined in Definition \ref{def:cut_off}.

\begin{definition}[\textbf{Matching Patterns}]\label{def:matching_patterns}
We consider the following match patterns, for $K\in\{S,B\}$, $\bar{K}\in\{S,B\}\backslash\{K\}$.
\begin{itemize}
    \item[(i)] Complete-matched on side $K$: $\delta^{K} = \underline{\lambda}^{K}$.
    \item[(ii)] Bottom-eliminated on side $K$: $\delta^{K}>\underline{\lambda}^{K}$. 
    %and $\tau^{\bar{K}}(\delta^{\bar{K}}|\delta^{\bar{K}})<\bar{\lambda}^{\bar{K}}$.
    \item[(iii)] Top-reserved on side $K$:  $\tau^{\bar{K}}(\delta^{\bar{K}}|\delta^{\bar{K}}) < \bar{\lambda}^{K}$.
    % \item[(iv)] Vertex matched on side $K$: $\tau^{\bar{K}}(\delta^{\bar{K}}|\delta^{\bar{K}}) = \bar{\lambda}^{K}$.
\end{itemize}
\end{definition}

Here, complete-matched on side $K$ means all participants on side $K$ are matched to (some or all of) participants on side $\bar{K}$. If the market is bottom-eliminated on side $K$, then the participants of types in $[\underline{\lambda}^{K}, \delta^{K})$ are not matched to any participants.
If the market is top-reserved on side $K$, then there is a group of types in the top subset $[\tau^{\bar{K}}(\delta^{\bar{K}}|\delta^{\bar{K}})$ that are matched to all the participants on side $\bar{K}$ whose types are in $[\delta^{\bar{K}}, \bar{\lambda}^{\bar{K}}]$, i.e., those are not bottom-eliminated on side $\bar{K}$.

% $$AAAA$$
% %
% Their preferences over the opponents of lower types, however, exhibit a asymmetric relation.
% % In this data market model, the seller side and the buyer side, however, are not symmetric with respect to the preferences over the matching.
% Specifically, a seller of $\lambda^{S}$ can provide data values $v(\lambda^{S},\lambda^{B}_{j})$ and $v(\lambda^{S},\lambda^{B}_{k})$ to buyer $j$ of $\lambda^{B}_{j}$ and buyer $k$ of $\lambda^{B}_{k}$, respectively.
% %
% In general, $v(\lambda^{S},\lambda^{B}_{j}) \neq v(\lambda^{S},\lambda^{B}_{k})$ if $\lambda^{B}_{j} = \lambda^{B}_{k}$.
% %
% However, the seller's valuation of her data is contained in her type $\lambda^{S}$ and is independent of the matched buyers' types. %
% Since the reward $M^{S}(\gamma(\lambda^{S}, \lambda^{B});\lambda^{S})$ is non-negative and the same data (cpoy) can be shared to multiple buyers, each seller prefers to be matched to buyers of as a wider range of types from low to high as possible.
% %
% The buyers, on the other hand, have to tradeoff the received value from the shared data and the rewards provided to each of the matched types.
%

%

Based on the cut-off rule, we can rewrite the utility function in (\ref{eq:utility}) as follows, with a slight abuse of notation:
\begin{equation}\label{eq:utility_cut_off}
    u^{K}(\lambda^{K}, \tau^{K}(\lambda^{K}|\delta^{K})|\sigma^{K}) = \int^{\bar{\lambda}^{\bar{K}}}_{\tau^{K}(\lambda^{K}|\delta^{K})} R^{K}(\lambda^{K},x|\sigma^{K}) f^{\bar{K}}(x) dx.
\end{equation}
%
%Clearly, $u^{K}(\lambda^{K}, \tilde{\lambda}^{\bar{K}}|\sigma^{K})$ is an increasing function of $\tilde{\lambda}^{\bar{K}}$.
%
%{eq:revenue} {eq:social_welfare}

The platform's revenue in (\ref{eq:revenue_v1}) can be rewritten, by setting $J^{K}(\underline{\lambda}^{K}|\sigma^{K}, \phi^{K})=0$ and integral by parts as follows:
\begin{equation}\label{eq:revenue_v2}
    \begin{aligned}
    Z_{R}(\bm{\sigma}, \bm{\phi}) = \sum_{K\in\{S,B\}}& \int_{\Lambda^{K}} \Big[u^{K}(x, \tau^{K}(x|\delta^{K})|\sigma^{K}) \\
    &- D^{K}(x) \frac{1-F^{K}(x) }{f^{K}(x)}\Big] f^{K}(x) dx.
    \end{aligned}
\end{equation}

Then, the virtual surpluses of the welfare and the revenue maximization of the participant of $\lambda^{K}$, respectively, are as follows: for $K\in\{S,B\}$,
\begin{equation}\label{eq:virtual_surplus_welfare}
    U^{K}_{W}(\lambda^{K},\tau^{K}(\lambda^{K}|\delta^{K})) = u^{K}(\lambda^{K}, \tau^{K}(\lambda^{K}|\delta^{K})|\sigma^{K}),  
\end{equation}
and
\begin{equation}\label{eq:virtual_surplus_revenuw}
    \begin{split}
        &U^{K}_{R}(\lambda^{K},\tau^{K}(\lambda^{K}|\delta^{K})) \\
        = & u^{K}(\lambda^{K}, \tau^{K}(\lambda^{K}|\delta^{K})|\sigma^{K}) - D^{K}(\lambda^{K}) \frac{1-F^{K}(\lambda^{K})}{f^{K}(\lambda^{K})}.
    \end{split}
\end{equation}

Suppose a participant has the type $\lambda^{K} \in [\tau^{\bar{K}}(\lambda^{\bar{K}}), \bar{\lambda}^{K}]$, i.e., she is matched to participants from $\bar{K}$ of type $\lambda^{\bar{K}}$ and other participants whose types are above $\lambda^{\bar{K}}$.
Hence, the marginal effect of this participant on the utility of the participants from $\bar{K}$ of $\lambda^{\bar{K}}$ can be modeled as
\begin{equation}\label{eq:marginal_utility_S}
\begin{split}
\omega^{\bar{K}}(\lambda^{K})=& \gamma^{K}(\lambda^{K})f^{K}(\lambda^{K}).
    % \omega^{K}(\lambda^{K}, \lambda^{\bar{K}})=& \frac{\partial u^{\bar{K}}( \lambda^{\bar{K}}, x|\sigma^{K})  }{ \partial x  }\Big|_{x= \lambda^{K}}.
\end{split}
    % \omega^{K}(\lambda^{K}, \lambda^{\bar{K}})\equiv \frac{\partial u^{\bar{K}}( \lambda^{\bar{K}}, x)  }{ \partial x  }\Big|_{x= \lambda^{K}} = -R^{\bar{K}}(\lambda^{\bar{K}},\lambda^{K})f^{K}(\lambda^{K}).
\end{equation}
% and
% \begin{equation}\label{eq:marginal_utility_R}
%     \begin{split}
%         \omega^{B}(\lambda^{})
%     \end{split}
% \end{equation}
%
%
% Also, the direct marginal effect (no through the matched participants from $\bar{K}$) of her on her utility can be modeled as
% %
% \begin{equation}\label{eq:marginal_utility}
%     \begin{aligned}
%     \frac{\partial u^{K}(x, \tau^{K}(\lambda^{K}|\delta^{K})|\sigma^{K})  }{\partial x  }\Big|_{x = \lambda^{K}} = &\int^{\bar{\lambda}^{\bar{K}}}_{\tau^{K}(\lambda^{K}|\delta^{K})} \frac{\partial}{\partial x} R^{K}(x,y)dy. 
%     \end{aligned}
% \end{equation}
%
The direct marginal effect of this participant (not from influencing the matched participants from $\bar{K}$ through the matching rule)
%
%(not indirect effect from the the matched participants from $\bar{K}$ through participant $i$'s report) 
%
on the social welfare defined in (\ref{eq:social_welfare}) is proportional to the following, for some arbitrarily fixed non-zero $\lambda^{\bar{K}}_{\Delta}\in \Lambda^{\bar{K}}$,
\begin{equation}\label{eq:margianl_social_welfare}
\begin{split}
\theta^{K}_{W}(\lambda^{K}, \lambda^{\bar{K}}_{\Delta}) =& -\frac{\partial }{\partial x} U^{K}_{W}(\lambda^{K},x)f^{K}(\lambda^{K})\Big|_{x=\lambda^{\bar{K}}_{\Delta}}.
%   \theta^{K}_{W}(\lambda^{K}) =&  U^{K}_{W}(\lambda^{K},\lambda^{\bar{K}}_{\Delta})f^{K}(\lambda^{K}).
   %=& u^{K}(\lambda^{K}, \lambda^{\bar{K}}_{\Delta}|\sigma^{K})
    % \theta^{K}_{W}(\lambda^{K}, \tau^{K}(\lambda^{K})) =& \frac{\partial}{\partial x} U^{K}_{W}(\lambda^{K},x)\Big|_{x = \lambda^{\bar{K}}_{\Delta}}\\ =&\frac{\partial}{\partial x} u^{K}(\lambda^{K}, x|\delta^{K})|\sigma^{K})\Big|_{x = \lambda^{\bar{K}}_{\Delta}}.
\end{split}
    % \theta^{K}_{W}(\lambda^{K}) =  u^{K}(\lambda^{K}, \tau^{K}(\lambda^{K}|\delta^{K})|\sigma^{K})f^{K}(\lambda^{K}).
\end{equation}
%
%
% We can re-write the platform's revenue in (\ref{eq:revenue_v1}), by setting $J^{K}(\underline{\lambda}^{K}|\sigma^{K}, \phi^{K})\Big]f^{K}(x)=0$, through integral by parts as follows:
% %
% \begin{equation}\label{eq:revenue_v2}
%     \begin{aligned}
%     Z_{W}(\bm{\sigma}, \bm{\phi}) =& \sum_{K\in\{S,B\}} \int_{\Lambda^{K}} \Big[u^{K}(x, \tau^{K}(x|\delta^{K})) \\
%     &- D^{K}(x) \frac{1-F^{K}(x) }{f^{K}(x)}\Big] f^{K}(x) dx.
%     \end{aligned}
% \end{equation}
% %
Similarly, the direct marginal effect of participant $i$ on the platform's revenue defined in (\ref{eq:revenue}) is proportional to the following, for some arbitrary fixed $\lambda^{\bar{K}}_{\Delta}\in \Lambda^{\bar{K}}$,
\begin{equation}\label{eq:marginal_revenue}
    \begin{split}
        \theta^{K}_{R}(\lambda^{K},\lambda^{\bar{K}}_{\Delta})=-\frac{\partial}{\partial x}  U^{K}_{R}(\lambda^{K}, x)f^{K}(\lambda^{K})\Big|_{x=\lambda^{\bar{K}}_{\Delta}}.
        % \\ =&\frac{\partial}{\partial x} \big[u^{K}(\lambda^{K}, x)\\
        % &- D^{K}(\lambda^{K}) \frac{1-F^{K}(\lambda^{K}) }{f^{K}(\lambda^{K})}\big]\Big|_{x = \tau^{K}(\lambda^{K}|\delta^{K})} f^{K}(\lambda^{K}).
    \end{split}
\end{equation}
We have the following assumption.
\begin{assumption}\label{assp:regularity}
For $K\in\{S,B\}$, $Y\in \{W,R\}$, $\frac{\theta^{K}_{Y}(\lambda^{K},\lambda^{\bar{K}}_{\Delta}) }{ \omega^{\bar{K}}(\lambda^{K})}$ is a strictly increasing function of $\lambda^{K}$, for any fixed non-zero $\lambda^{\bar{K}}_{\Delta}\in \Lambda^{\bar{K}}$.
\end{assumption}

In Assumption \ref{assp:regularity}, the term $\frac{\theta^{K}_{Y}(\lambda^{K},\lambda^{\bar{K}}_{\Delta}) }{ \omega^{\bar{K}}(\lambda^{K})}$ is the ratio of a participant of $\lambda^{K}$'s direct marginal contribution to $Z_{Y}$, $Y\in\{W,R\}$, and her marginal contribution to each of her matched participants from $\bar{K}$.
Assumption \ref{assp:regularity} requires that the direct marginal contribution of a participant of $\lambda^{K}$ to the social welfare or the platform's revenue changes faster than her marginal contribution to any of her matched opponents from $\bar{K}$ through the matching rule.

Define, for $K\in \{S,B\}$, $Y\in\{W,R\}$,
\begin{equation}\label{eq:joint_marginal}
    \eta^{K}_{Y}(\lambda^{K}, \lambda^{\bar{K}}) = \theta^{K}_{Y}(\lambda^{K}, \lambda^{\bar{K}}) + \theta_{Y}^{\bar{K}}(\lambda^{\bar{K}}, \lambda^{K}).
\end{equation}
Here, $\eta^{K}_{Y}(\lambda^{K}, \lambda^{\bar{K}})$ is the joint marginal effect of matching $\lambda^{K}$ to $\lambda^{\bar{K}}$ to the social welfare ($Y=W$) or platform's revenue ($Y=R$) if they are matched. Here, $\eta^{K}_{Y}$ coincides with the contribution $C$ in Assumption \ref{assp:order_of_preference}. 

The following proposition shows the uniqueness of the cut-off functions.

\begin{proposition}\label{prop:uniqueness}
Suppose Assumptions \ref{assp:utility} and \ref{assp:regularity} hold. If $\eta^{K}_{Y}(\underline{\lambda}^{K}, \underline{\lambda}^{\bar{K}})< 0$, for $Y\in\{W,R\}$, $K\in\{S,B\}$, 
%and the market is not complete-matched on both sides, 
then the cut-off function $\tau^{K}$ specifies a unique matched lowest type for each type on side $K$, i.e., $\tau^{K}(\lambda^{K}|\delta^{K})\neq \tau^{K}(\tilde{\lambda}^{K}|\delta^{K})$, for any two different $\lambda^{K}$, $\tilde{\lambda}^{K}\in [\delta^{K}, \bar{\lambda}^{K}]$.
\end{proposition}

\proof
See Appendix \ref{app:prop_uniqueness}.
\endproof

The following proposition shows the conditions under which the market exhibits certain matching patterns. 

\begin{proposition}\label{prop:match_patterns}
Suppose Assumptions \ref{assp:utility} and \ref{assp:regularity} hold. Then, the followings are ture.
\begin{itemize}
    \item[(i)] If $\eta^{K}_{Y}(\underline{\lambda}^{K}, \underline{\lambda}^{\bar{K}})\geq 0 $, then the market is complete-matched on both sides of the market, for $Y\in\{W,R\}$, $K\in\{S,B\}$.
    \item[(ii)]If $\eta^{K}_{Y}(\underline{\lambda}^{K}, \underline{\lambda}^{\bar{K}})< 0$ and $\eta^{K}_{Y}(\bar{\lambda}^{K}, \underline{\lambda}^{\bar{K}})> 0$, then the market is top-reserved on side $K$ and
    complete-matched on side $\bar{K}$. Additionally $\tau^{K}(\lambda^{K}|\delta^{K}) = \delta^{\bar{K}}$ for all $\lambda^{K}\in[\tau^{\bar{K}}|\delta^{\bar{K}}, \bar{\lambda}^{K}]$.
    \item[(iii)] If $\eta^{K}_{Y}(\underline{\lambda}^{K}, \underline{\lambda}^{\bar{K}})< 0$ and $\eta^{K}_{Y}(\bar{\lambda}^{K}, \underline{\lambda}^{\bar{K}})<0$, then the market is not top-reserved on side $K$ and bottom-eliminated on side $\bar{K}$.
    \item[(iv)] If $\eta^{K}_{Y}(\underline{\lambda}^{K}, \underline{\lambda}^{\bar{K}})< 0$ and $\eta^{K}_{Y}(\bar{\lambda}^{K}, \underline{\lambda}^{\bar{K}})=0$, then the market is not top-reserved on side $K$ and complete-matched on side $\bar{K}$.
\end{itemize}
\end{proposition}
\proof

See Appendix \ref{app:prop_match_patterns}.

\endproof

Without loss of generality, we refer to $K$ as the seller and $\bar{K}$ as the buyer and focus on the social welfare maximization, i.e., $Y=W$.
In Proposition \ref{prop:match_patterns}, $\eta^{S}_{W}(\underline{\lambda}^{S}, \underline{\lambda}^{B})$ is the joint marginal effect of matching the sellers' lowest type $\underline{\lambda}^{S}$ to the buyers lowest type $\underline{\lambda}^{B}$.
It $\eta^{S}_{W}(\underline{\lambda}^{S}, \underline{\lambda}^{B})\geq 0$, then $\eta^{S}_{W}(\lambda^{S}, \lambda^{B})\geq0$ for all $\lambda^{S}\in \Lambda^{S}$ and $\lambda^{B}\in\Lambda^{B}$ due to the monotonicity in Assumption  \ref{assp:utility}.
If $\eta^{S}_{W}(\underline{\lambda}^{S}, \underline{\lambda}^{B})< 0$, then the lowest types should not be matched because they contribute negatively to the social welfare.
Here, $\eta^{S}_{W}(\bar{\lambda}^{S}, \underline{\lambda}^{B})$ measures the joint marginal effect of matching a seller of the highest type $\bar{\lambda}^{S}$ to a buyer of the lowest type $\underline{\lambda}^{B}$.
Together with $\eta^{S}_{W}(\underline{\lambda}^{S}, \underline{\lambda}^{B})< 0$, $\eta^{S}_{W}(\bar{\lambda}^{S}, \underline{\lambda}^{B})>0$ means that buyers of the lowest type can generate positive contribution to the social welfare, which implies that buyers of all kinds of types are matched to sellers and $\tau^{B}(\underline{\lambda}^{B}|\delta^{B})<\bar{\lambda}^{S}$, i.e., top-reserved, due to the non-increasing property of $\tau^{B}$.
On the other hand, $\eta^{S}_{W}(\underline{\lambda}^{S}, \underline{\lambda}^{B})< 0$ and $\eta^{S}_{W}(\bar{\lambda}^{S}, \underline{\lambda}^{B})<0$ imply that the buyers of the lowest type should be eliminated from being matched. This exhibits the bottom-eliminated on the buyer side, i.e., $\delta^{B}>\underline{\lambda}^{B}$.
Hence, $\tau^{B}(\delta^{B}|\delta^{B})<\tau^{B}(\underline{\lambda}^{B}|\delta^{B})$. 
Since $\underline{\lambda}^{B}$ is excluded (and not matched to $\bar{\lambda}^{S}$), $\tau^{B}(\delta^{B}|\delta^{B})= \bar{\lambda}^{S}$, i.e., the buyers of lowest matched type is matched only to the sellers of the highest type.
From the reciprocal relationship, measuring $\eta^{B}_{W}$ on $\underline{\lambda}^{S}$, $\underline{\lambda}^{B}$, and $\bar{\lambda}^{S}$ can complete the matching patterns on both sides. 
Corollary \ref{corollary:tau} directly follows Propositions \ref{prop:uniqueness} and \ref{prop:match_patterns}.

\begin{corollary}\label{corollary:tau}
Suppose Assumptions \ref{assp:utility} and \ref{assp:regularity} hold. Then, the followings hold.
\begin{itemize}
    \item[(i)] If  $\eta^{K}_{Y}(\underline{\lambda}^{K}, \underline{\lambda}^{\bar{K}})<0$, then $\eta^{K}_{Y}(\lambda^{K}, \tau^{K}(\lambda^{K}|\delta^{K}))=0$.
    \item[(ii)] If $\eta^{K}_{Y}(\underline{\lambda}^{K}, \underline{\lambda}^{\bar{K}})< 0$ and $\eta^{K}_{Y}(\bar{\lambda}^{K}, \underline{\lambda}^{\bar{K}})<0$, then $\tau^{\bar{K}}(\delta^{\bar{K}}|\delta^{\bar{K}}) = \bar{\lambda}^{K}$.
    \item[(iii)]  If $\eta^{K}_{Y}(\underline{\lambda}^{K}, \underline{\lambda}^{\bar{K}})< 0$ and $\eta^{K}_{Y}(\bar{\lambda}^{K}, \underline{\lambda}^{\bar{K}})=0$, then $\tau^{\bar{K}}(\underline{\lambda}^{\bar{K}}|\underline{\lambda}^{\bar{K}}) = \bar{\lambda}^{K}$.
\end{itemize}

\end{corollary}

\begin{figure*}[htbp]
\centering
\subfloat[]{\label{welfare_fig}\includegraphics[width=0.3\textwidth]{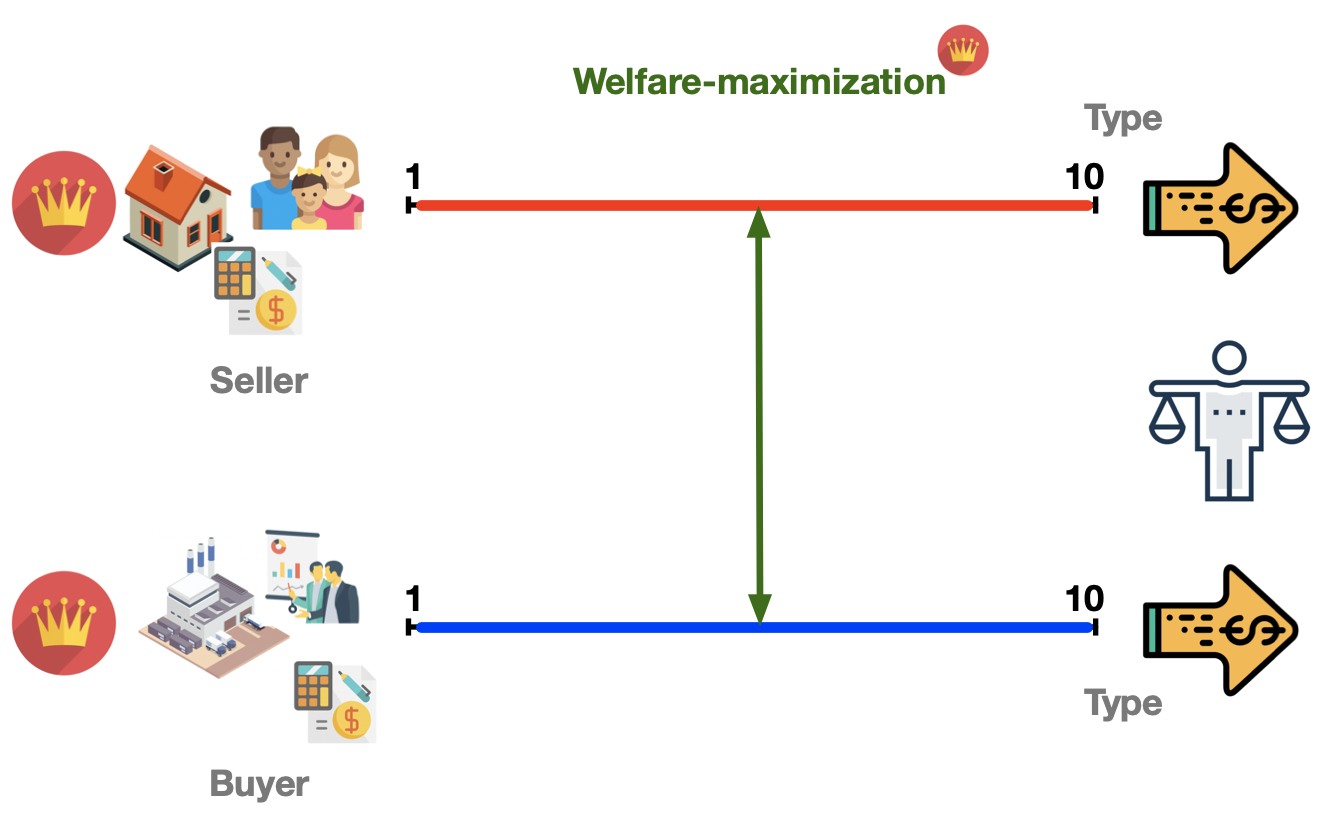}}
\subfloat[]{\label{revenue_fig}\includegraphics[width=0.3\textwidth]{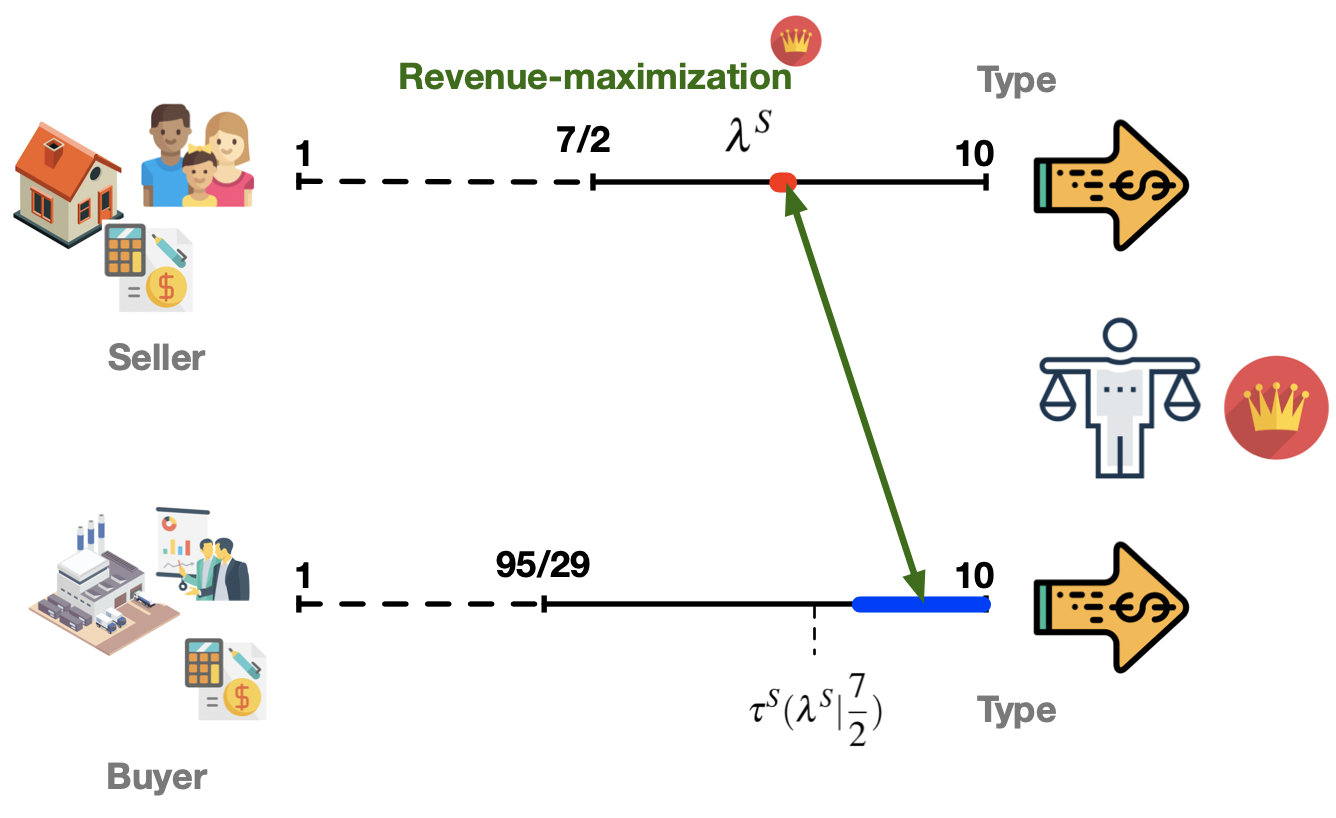}}
\subfloat[]{\label{welfare_opt}\includegraphics[width=0.19\textwidth]{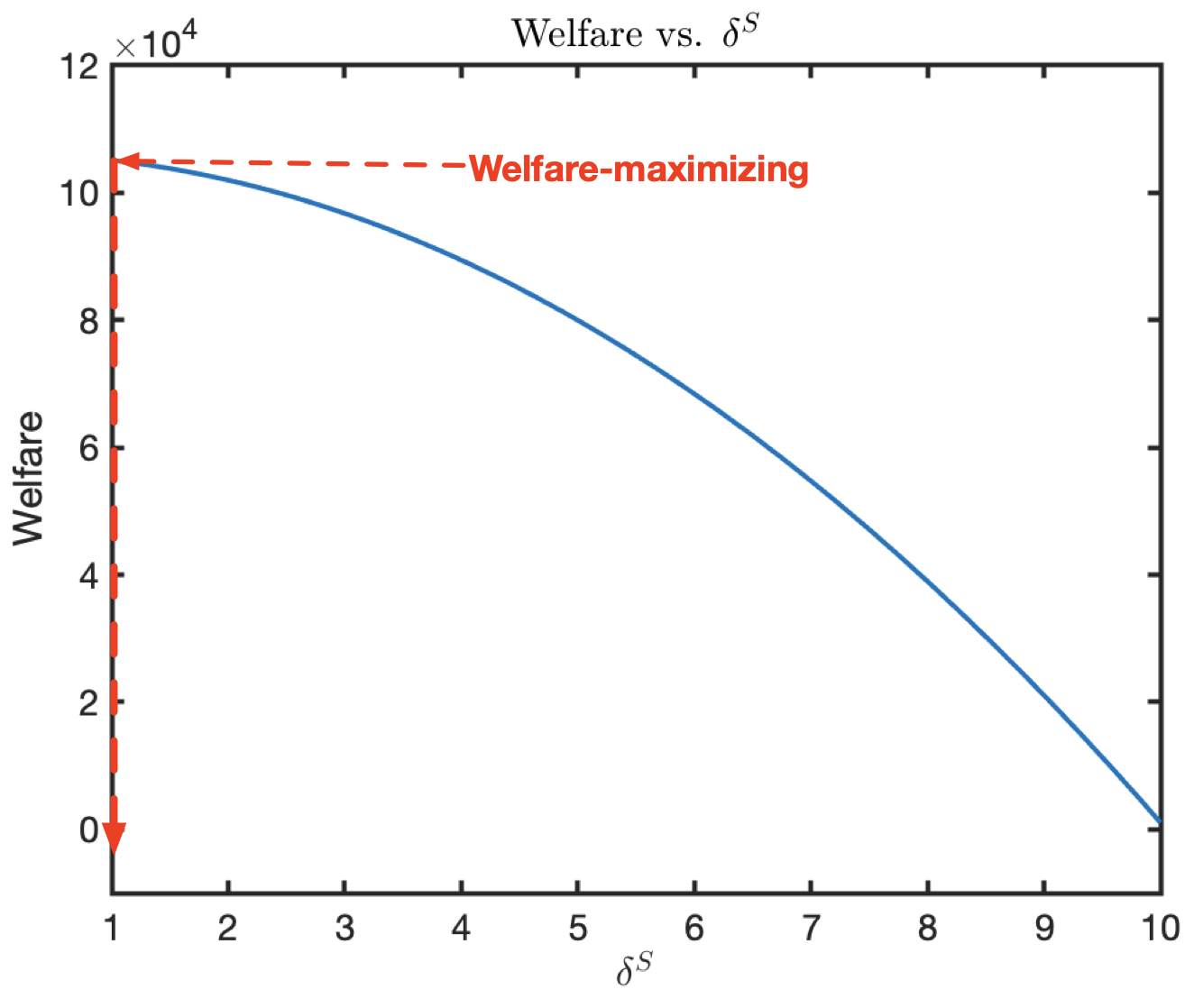}}
\subfloat[]{\label{revenue_opt}\includegraphics[width=0.20\textwidth]{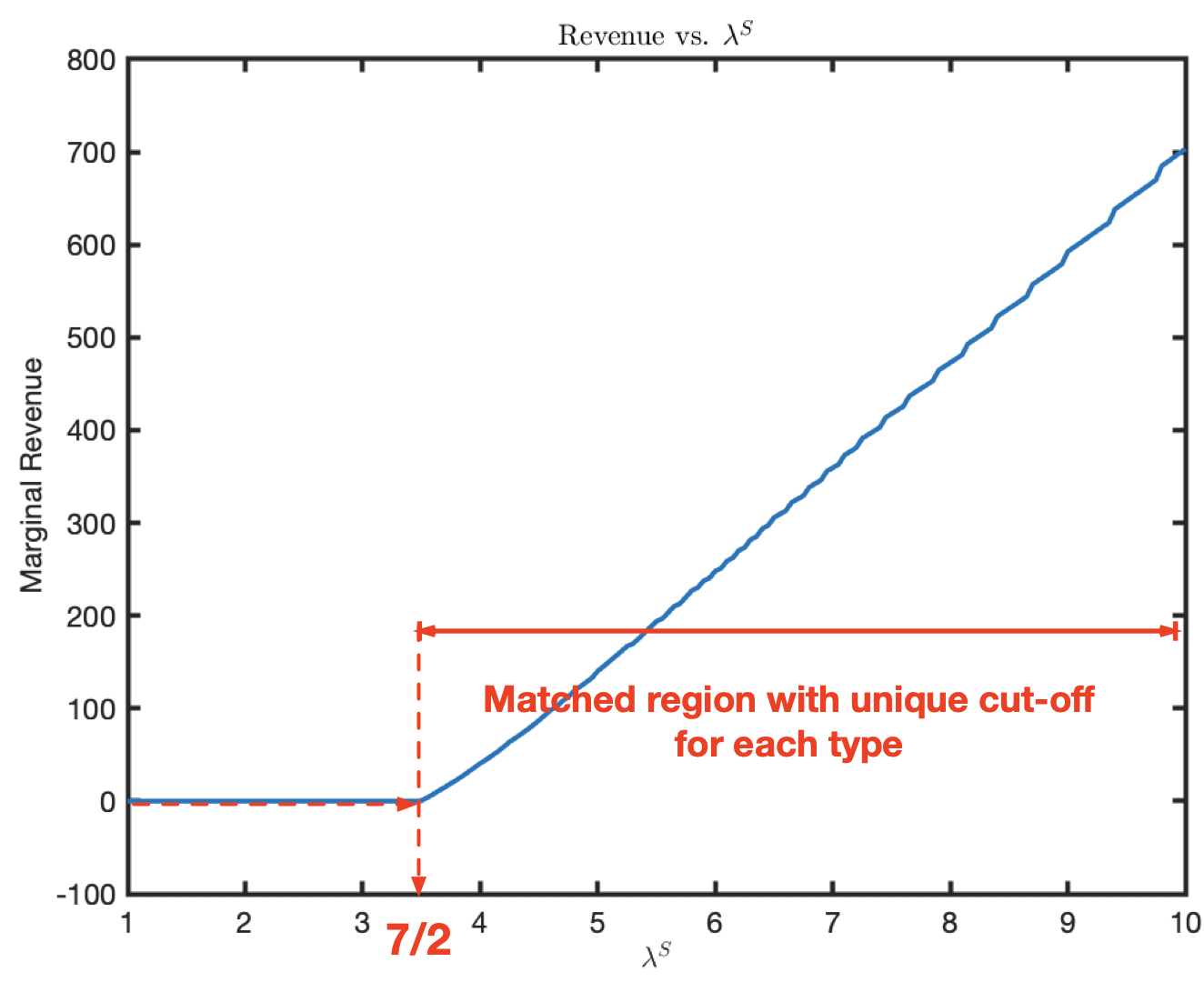}}%
\caption{(a): Welfare-maximizing matching rule. $\delta^{K}=\underline{\lambda}^{K}$ and $\tau^{K}(\underline{\lambda}^{K}|\delta^{K}) = \underline{\lambda}^{\bar{K}}$.
i.e., each data seller is matched to all data buyers. (b): Revenue-maximizing matching rule. Each data seller of $\lambda^S$ is matched to all data buyers of $\lambda^B\in[\tau^{S}(\lambda^{S}|\frac{40}{9}), 10]$, where $\tau^{S}(\lambda^{S}|\frac{7}{2}) = \frac{10\lambda^{S}-5 }{4\lambda^{S} -11}$. (c): Welfare vs. $\delta^K$. When $\delta^K = 1$, the welfare reaches its maximum, i.e., $\tau^{\bar{K}}(\delta^{\bar{K}}|\delta^{\bar{K}}) = \underline{\lambda}^{K}$. (d): Marginal Revenue vs. $\lambda^{S}$. The platform has a non-negative marginal revenue for $ \lambda^{S}\geq \frac{7}{2}$. $\tau^{S}(\lambda^{S}|\frac{7}{2}) = \frac{10\lambda^{S}-5 }{4\lambda^{S} -11}$ is optimal and specifies a unique cut-off for each $\lambda^{S}\in[\frac{7}{2}, 10]$.}
\label{fig:myfig}
\end{figure*}

\subsection{Numerical Analysis}\label{Sect_5}
%\textcolor{red}{$h^L$ is a set of index??? No.}

% In this section, we present a numerical example for the matching to illustrate the optimal matching rules for maximizations of the social welfare and the platform's revenue.
% %
% Consider $\gamma^{S}(\lambda^{S}) = \lambda^{S}$, $\gamma^{B}(\lambda^{B})=\frac{1}{2}\lambda^{B}$, $M^{K}(r, \lambda^{K})=C^{K}\lambda^{K}r$, with $C^{S}=1$ and $C^{B}=\frac{1}{2}$, for $K\in\{S,B\}$.
% %
% Hence, $R^{S}(\lambda^{S},\lambda^{B}|\sigma^{S}) =\lambda^{S}\lambda^{B}$ and $R^{B}(\lambda^{B}, \lambda^{S}|\sigma^{B}) = \frac{1}{4}\lambda^{S}\lambda^{B}$.
% %
% Suppose $\lambda^{K}$ is uniformly distributed over $\Lambda^{K}=[\underline{\lambda}, \bar{\lambda}]$, for $K\in\{S,B\}$.

% \textcolor{red}{bunching: many to one}

% \textcolor{red}{$$AAAAA$$}

In this section, we present a numerical example for the matching to illustrate the optimal matching rules for the social welfare and the platform's revenue maximization.
Fig. \ref{fig:myfig} shows the numerical results.
Consider the case when $\Lambda^{K}=[1,10]$,
$\gamma^{S}(\lambda^{S}) = \lambda^{S}$, $\gamma^{B}(\lambda^{B})=\frac{1}{2}\lambda^{B}$, $M^{K}(r, \lambda^{K})=C^{K}\lambda^{K}r$, with $C^{S}=1$ and $C^{B}=\frac{1}{2}$, for $K\in\{S,B\}$.
Let $R^{S}(\lambda^{S},\lambda^{B}|\sigma^{S}) =\frac{1}{2}\lambda^{S}\lambda^{B}$ and $R^{B}(\lambda^{B}, \lambda^{S}|\sigma^{B}) = \frac{1}{2}\lambda^{B}(\lambda^{S}-\frac{1}{2})$.
% Consider the case when $\Lambda^{K}=[1,10]$,
% $\gamma^{S}(\lambda^{S}) = \lambda^{S}$, $\gamma^{B}(\lambda^{B})=\frac{1}{2}\lambda^{B}$, $M^{K}(r, \lambda^{K})=C^{K}\lambda^{K}r$, with $C^{S}=1$ and $C^{B}=\frac{1}{2}$, for $K\in\{S,B\}$.
% %
% Hence, $R^{S}(\lambda^{S},\lambda^{B}|\sigma^{S}) =\frac{1}{2}\lambda^{S}\lambda^{B}$ and $R^{B}(\lambda^{B}, \lambda^{S}|\sigma^{B}) = \frac{1}{2}\lambda^{B}(\lambda^{S}-\frac{1}{2}\lambda^{B})$.
%
Suppose the type $\lambda^{K}$ is uniformly distributed over $\Lambda^{K}$, for $K\in\{S,B\}$, i.e., $F^{K}(\lambda^{K})=\frac{\lambda^{K}-1}{9}$, and $f^{K}(\lambda^{K}) = \frac{1}{9}$.
Then, $\omega^{B}(\lambda^{S})= \frac{1}{9}\lambda^{S}$, $\omega^{S}(\lambda^{B})=\frac{1}{18}\lambda^{B}$, $\theta^{S}_{W}(\lambda^{S}, \lambda^{B}_{\Delta})=\frac{1}{18}\lambda^{B}_{\Delta}\lambda^{S}$, $\theta^{B}_{W}(\lambda^{B}, \lambda^{S}_{\Delta}) =\frac{1}{18}\lambda^{B}(\lambda^{S}_{\Delta}-\frac{1}{2})$, $\theta^{S}_{R}(\lambda^{S},\lambda^{B}_{\Delta})= \frac{1}{9}\lambda^{B}_{\Delta}(\lambda^{S}-5)$, $\theta^{B}_{R}(\lambda^{B}, \lambda^{S}_{\Delta}) = \frac{1}{9}(\lambda^{B}-5)(\lambda^{S}-\frac{1}{2})$.
%
% $$AAAAA$$
% Then,   , , $\theta^{B}_{R}(\lambda^{B},\lambda^{S}_{\Delta})= \frac{1}{18}(2\lambda^{S}_{\Delta}\lambda^{B} - \frac{3}{2}\{\lambda^{B}\}^{2}+10\lambda^{B} - 10\lambda^{S}_{\Delta})$.
% %
It is straightforward to verify that Assumption \ref{assp:regularity} holds on both sides of the market for both the social welfare and the revenue maximization.
For $K\in\{S,B\}$, 
$$
\eta^{K}_{W}(\lambda^{K}, \lambda^{\bar{K}}) = \frac{1}{36}\lambda^{B}(4\lambda^{S} -1), %\frac{1}{18}\lambda^{B}\lambda^{S} + \frac{1}{18}\lambda^{B}(\lambda^{S}-\frac{1}{2}\lambda^{B}),
$$
$$
\eta^{K}_{R}(\lambda^{K}, \lambda^{\bar{K}}) = \frac{1}{9}(2\lambda^{B}\lambda^{S}-5\lambda^{S} -\frac{11}{2}\lambda^{B}+\frac{5}{2}). %\frac{2}{9}\lambda^{B}\lambda^{S} +\frac{5}{9}(\lambda^{B}-\lambda^{S}-1) + \frac{1}{12}\{\lambda^{B}\}^{2}.
% \frac{1}{9}\lambda^{B}(\lambda^{S}-5) + \frac{1}{18}(2\lambda^{S}\lambda^{B} - \frac{3}{2}\{\lambda^{B}\}^{2} + 10\lambda^{B}-10\lambda^{S}).
$$
Since $\eta^{K}_{W}(\lambda^{K}, \lambda^{\bar{K}})>0$ for all $\lambda^{K}\in \Lambda^{K}$ and $\lambda^{\bar{K}}\in \Lambda^{\bar{K}}$, social welfare maximizing matching rule matches all sellers to all buyers and reciprocally all buyers to all sellers.
For revenue maximization, since $\eta^{K}_{R}(1,1)<0$, for $K\in\{S,B\}$, Proposition \ref{prop:uniqueness} implies that the matching rule sets a unique $\tau^{K}(\lambda^{K}|\delta^{K})$ for each $\lambda^{K}\in [\delta^{K},\bar{\lambda}^{K}]$.
On the seller side, $\eta^{S}_{R}(10,1) = -\frac{33}{9}<0$. 
One the buyer side, $\eta^{B}_{R}(10,1) = -\frac{75}{2}<0$.
Hence, Proposition \ref{prop:match_patterns} implies that the market is bottom-eliminated and not top-reserved on both sides.
From Corollary \ref{corollary:tau}, we have
$$
\tau^{S}(\lambda^{S}|\delta^{S}) = \frac{10\lambda^{S}-5 }{4\lambda^{S} -11},
$$
with $\delta^{S}=\frac{7}{2}$, and
$$
\tau^{B}(\lambda^{B}|\delta^{B}) = \frac{11\lambda^{B}-5}{4\lambda^{B}-10},
$$
with $\delta^{B}=\frac{95}{29}$.

\section{Conclusion}\label{Sect_6}

In this paper, we have proposed a two-sided matching market framework for IoT data trading as a sustainable pricing model that incentivizes both the data sellers and the data buyers. A monopolist platform has been introduced to match the data owners and the data buyers based on their private information. We have established a quantitative framework to model the IoT data trading market for the welfare and the revenue maximization. This work has characterized a class of feasible cut-off matching rules. Under mild assumptions of the utility functions and the distributions of the valuations, there exist optimal cut-off matching rules that maximize the welfare and the revenue. We have provided numerical example to understand the 
matching patterns of participants from both sides of the market.
\bibliographystyle{IEEEtran}
\bibliography{IEEEabrv,TAO_LCSL}

%%%%%%%%%%%%%%%%%%%
\appendix

\subsection{Proof of Proposition \ref{prop:uniqueness}}\label{app:prop_uniqueness}

Expanding $\theta^{K}_{W}$ and $\theta^{K}_{R}$ as follows:
$$
\begin{aligned}
\theta^{K}_{W}(\lambda^{K}, \lambda^{\bar{K}}) =& -\frac{\partial }{\partial x} U^{K}_{W}(\lambda^{K},x)f^{K}(\lambda^{K})\Big|_{x=\lambda^{\bar{K}}}\\
=& R^{K}(\lambda^{K}, \lambda^{\bar{K}}|\sigma^{K})f^{\bar{K}}(\lambda^{\bar{K}})f^{K}(\lambda^{K}),
\end{aligned}
$$
and
$$
\begin{aligned}
\theta^{K}_{R}(\lambda^{K},\lambda^{\bar{K}}_{\Delta})=&-\frac{\partial}{\partial x}  U^{K}_{R}(\lambda^{K}, x)f^{K}(\lambda^{K})\Big|_{x=\lambda^{\bar{K}}_{\Delta}}\\
=&R^{K}(\lambda^{K}, \lambda^{\bar{K}}|\sigma^{K})f^{\bar{K}}(\lambda^{\bar{K}})f^{K}(\lambda^{K})\\
&+ \frac{\partial}{\partial x} R^{K}(x, \lambda^{\bar{K}}|\sigma^{K})f^{\bar{K}}(\lambda^{\bar{K}})(1-F^{K}(\lambda^{K})).
\end{aligned}
$$
%

% It is straightforward to see that $\theta^{S}_{W}(\lambda^{S}, \lambda^{B})\geq 0$ for all $\lambda^{S}\in \Lambda^{S}$, $\lambda^{B}\in \Lambda^{B}$, $Y\in\{W,R\}$.
%

Define,
$$
\begin{aligned}
\rho^{K}_{W}(\lambda^{K}, \lambda^{\bar{K}}) =R^{K}(\lambda^{K}, \lambda^{\bar{K}}|\sigma^{K})f^{K}(\lambda^{\bar{K}}), 
\end{aligned}
$$
and
$$
\begin{aligned}
\rho^{K}_{R}(\lambda^{K}, \lambda^{\bar{K}}) =&R^{K}(\lambda^{K}, \lambda^{\bar{K}}|\sigma^{K})f^{K}(\lambda^{\bar{K}}) \\
&+  \frac{\partial}{\partial x} R^{K}(x, \lambda^{\bar{K}}|\sigma^{K})(1-F^{K}(\lambda^{K})),
\end{aligned}
$$
such that $\theta^{K}_{Y}(\lambda^{K}, \lambda^{\bar{K}})=\rho^{K}_{Y}(\lambda^{K}, \lambda^{\bar{K}})f^{\bar{K}}(\lambda^{\bar{K}})$.

If $\rho^{K}_{Y}(\underline{\lambda}^{K}, \underline{\lambda}^{\bar{K}})<0$, for all $K\in\{S,B\}$, $Y\in\{W,R\}$, then $\eta^{K}_{Y}(\underline{\lambda}^{K}, \underline{\lambda}^{\bar{K}})<0$.
Let $\tau^{K}(|\delta^{K})$ be the matching rule that satisfies the uniqueness property in Proposition \ref{prop:uniqueness}.
Let $\hat{\tau}^{K}(|\hat{\delta}^{K})$ with threshold $\hat{\delta}^{K}$ be the cut-off function that does not specify unique matched cut-off types for different types.
Hence, participants from side $K$ are excluded from matching if $\lambda^{K}<\hat{\delta}^{K}$. Otherwise, participants from side $K$ are matched to participants from side $\bar{K}$ whose types are above $\hat{\delta}^{\bar{K}}$.

Let $g^{K}_{Y} = \inf\{\lambda^{K}\in\Lambda^{K}: \rho^{K}_{Y} \geq 0\}$. 
Suppose the platform adopts the unique matching rule $\tau^{K}$ that matches each participant of type $\lambda^{K}\geq \hat{\delta}^{K}$ to the same set as the non-unique matching rule $\hat{\tau}^{K}$ and matches each participant of type $\lambda^{K}\in[g^{K}_{Y},  \hat{\delta}^{K}]$ to the set $[\tilde{\lambda}^{\bar{K}}, \bar{\lambda}^{\bar{K}}]$, where $\tilde{\lambda}^{\bar{K}} = \max\{g^{K}_{Y}, \hat{\delta}^{K}\}$.
Then, the platform can increase his payoff (utilities or revenues) by switching $\hat{\tau}^{K}$ to $\tau^{K}$.

Now, suppose $\hat{\delta}^{K}<g^{K}_{Y}$ for $K\in\{S,B\}$.
Let $\tau^{K}_{\Omega}$ be defined as follows:
$$
\tau^{K}_{\Omega}(\lambda^{K}) \begin{cases}
[\hat{\delta}^{\bar{K}}, \bar{\lambda}^{\bar{K}}] & \text{ if, } [g^{K}_{Y}, \bar{v}^{K}]\\
[g^{\bar{K}}_{Y}, \bar{\lambda}^{K}] & \text{ if, }  [\hat{\delta}^{K}, g^{K}_{Y}]\\
\emptyset & \text{ if, } [\underline{\lambda}^{K}, \hat{\delta}^{K}].
\end{cases}
$$
By adopting $\tau^{K}_{\Omega}$, the platform can improve his payoff than using the original $\hat{\tau}^{K}$ with $\bar{\delta}^{K}$ because all types that lead to negative $\rho^{K}_{Y}$ are eliminated, for $K\in\{S,B\}$, $Y\in\{S,B\}$.

Next, suppose $\hat{\delta} = g^{K}_{Y}$, for $K\in\{S,B\}$ and $\hat{\delta}^{\bar{K}}\leq g^{\bar{K}}_{Y}$, for $\bar{K}\in \{S,B\}\backslash{\{K\}}$, $Y\in\{W,R\}$.
Let $\tilde{\delta}^{K}+\epsilon = g^{K}_{Y}$ with sufficiently small $\epsilon>0$.
Define
$$
\tau^{K}_{\Omega}(\lambda^{K}) \begin{cases}
[\hat{\delta}^{\bar{K}}, \bar{\lambda}^{\bar{K}}] & \text{ if, } [g^{K}_{Y}, \bar{v}^{K}]\\
[g^{\bar{K}}_{Y}, \bar{\lambda}^{K}] & \text{ if, }  [\tilde{\delta}^{K}, g^{K}_{Y}]\\
\emptyset & \text{ if, } [\underline{\lambda}^{K}, \tilde{\delta}^{K}].
\end{cases}
$$
Here, participants of type $\lambda^{K}\in [\tilde{\delta}^{K}, g^{K}_{Y}]$ contribute negatively to the platform's payoff but they are sufficiently small such that they do not offset the positive contribution from participants of types in $[g^{K}_{Y}, \bar{v}^{K}]$.
Hence, the platform can increase his payoff by using $\tau^{K}_{\Omega}$ to replace the original $\hat{\tau}^{K}$ with $\hat{\delta}^{K}$.

%Next, consider $\rho^{\bar{K}}_{Y}(\underline{\lambda}^{\bar{K}}, \underline{\lambda}^{K})<0\leq \rho^{K}_{Y}(\underline{\lambda}^{K}, \underline{\lambda}^{\bar{K}})$.
%
% First, suppose $\eta^{K}_{Y}(\underline{\lambda}^{K}, \underline{\lambda}^{\bar{K}})\geq 0$, then the platform uses a matching rule $\tau^{K}_{\Gamma}(\lambda^{K})\geq \underline{v}^{\bar{K}}$ for some $\lambda^{K}\in \Lambda^{K}$. 
% %
% Let $\lambda^{K}_{a}$ be any type in $\Lambda^{K}$, such that $\tau^{K}$ is strictly decreasing at the right neighborhood of $\lambda^{K}_{a}$.
% %
% Hence, the marginal effect of reducing $\tau^{K}(\lambda^{K})$ below $\lambda^{\bar{K}}$ is $\eta^{K}_{Y}(\lambda^{K}, \lambda^{\bar{K}})$.
% %
% %
% Let $[\lambda^{K}_{\ell}, \lambda^{K}_{h}]\subset \Lambda^{K}$, in which $\tau^{K}(\lambda^{K}) = \lambda^{\bar{K}}$ for all $\lambda^{K}\in [\lambda^{K}_{\ell}, \lambda^{K}_{h}]$, i.e., $\tau^{K}$ specifies a constant cut-off type $\lambda^{\bar{K}}$, in which the marginal effect of the reduction of $\tau^{K}(\lambda^{K})$ below $\lambda^{\bar{K}}$ is $\int^{\lambda^{K}_{h}}_{\lambda^{K}_{\ell}}\eta^{K}_{Y}(x, \lambda^{\bar{K}})dx$.
% %
% %
% Under Assumption \ref{assp:regularity}, we have $\eta^{K}_{\Gamma}(\lambda^{K}, \lambda^{\bar{K}})>0$ for any $\lambda^{K}\in \Lambda^{K}$ and $\lambda^{\bar{K}}\in \Lambda^{\bar{K}}$, which implies that a complete-matched pattern for both sides is optimal.

Now, consider $\rho^{\bar{K}}_{Y}(\underline{\lambda}^{\bar{K}}, \underline{\lambda}^{K})<0\leq \rho^{K}_{Y}(\underline{\lambda}^{K}, \underline{\lambda}^{\bar{K}})$, $\eta^{K}_{Y}(\underline{\lambda}^{K}, \underline{\lambda}^{\bar{K}})< 0$ and the platform adopts a non-unique cut-off function.
First, suppose that the market is complete-matched on both sides, i.e., $\hat{\delta}^{K} = \underline{\lambda}^{K}$, $\tau^{\bar{K}}(\underline{\lambda}^{\bar{K}}) = \underline{\lambda}^{K}$.
Since $\eta^{K}_{Y}(\underline{\lambda}^{K}, \underline{\lambda}^{\bar{K}})< 0$, the platform's effort of increasing $\tau^{K}(\underline{\lambda}^{K}, )$ above $\underline{\lambda}^{\bar{K}}$ while keeping the cut-off type unchanged for all other types on side $K$ contributes to his payoff positively.
Hence, the platform can improve his payoff by using the unique cut-off by increasing $\tau^{K}$ at the right neighborhood of $\underline{\lambda}^{K}$ while keeping the cut-off type unchanged for all other types on side $K$.

Next, suppose that the original cut-off function eliminates some participants and match each of the participants of types above $\hat{\delta}^{K}$ to the same matched set.
As similar to the above analysis, optimal rule has to be in the case when $\hat{\delta}^{\bar{K}}< g^{\bar{K}}_{Y}$ and $\hat{\delta}^{K} = \underline{\lambda}^{K}$, and
$$
\int_{\Lambda^{K}}\eta^{K}_{Y}(x, \hat{\delta}^{\bar{K}})dx = 0
$$
%
% $$
% \begin{aligned}
% \theta^{\bar{K}}(\hat{\delta}^{\bar{K}}, \underline{\lambda}^{K}) = -\int_{\Lambda^{K}} \theta^{K}(x, \hat{\delta}^{\bar{K}})f^{K}(x)dx
% \end{aligned},
% $$
which requires that the total effect of an increase of the size of the matched set on $\bar{K}$ is zero. 
From Assumption \ref{assp:regularity}, we have there exists a $\lambda^{K}_{\Delta}\in \Lambda^{K}$ such that $\int^{\bar{\lambda}^{K}}_{\lambda^{K}_{\Delta}}\eta^{K}_{Y}(x, \hat{\delta}^{\bar{K}})dx>0$.
Thus, there exists a $\delta^{K}_{\Delta}<\hat{\omega}^{K}$ such that the platform can improve his payoff by using the following cut-off functions, instead of the original cut-off function:
$$
\tau^{K}_{\Omega}(\lambda^{K}) \begin{cases}
[\delta^{K}_{\Delta}, \bar{\lambda}^{\bar{K}}] & \text{ if, } [\lambda^{K}_{\Delta}, \bar{v}^{K}]\\
[\hat{\omega}^{K}, \bar{\lambda}^{K}] & \text{ if, }  [\underline{\lambda}^{K}, \lambda^{K}_{\Delta}].
\end{cases}
$$
Therefore, we can conclude that unique cut-off rule is optimal for the platform.

%%%%%%%%%%%%%%
\subsection{Proof of Proposition \ref{prop:match_patterns}}\label{app:prop_match_patterns}

% It is straightforward to verify that 
% $u^{K}(\lambda^{K}, \tau^{K}(\lambda^{K}|\delta^{K})|\sigma^{K})$ in (\ref{eq:utility_cut_off}) is decreasing in $\tau^{K}(\lambda^{K}|\delta^{K})$.
%

Let $\rho^{K}_{Y}$ be defined in Appendix \ref{app:prop_uniqueness},
such that $\theta^{K}_{Y}(\lambda^{K}, \lambda^{\bar{K}})=\rho^{K}_{Y}(\lambda^{K}, \lambda^{\bar{K}})f^{\bar{K}}(\lambda^{\bar{K}})$.

If $\rho^{K}_{Y}(\underline{\lambda}^{K}, \underline{\lambda}^{\bar{K}}) \geq 0$, for all $K\in\{S,B\}$, $Y\in\{W,R\}$, then $\eta^{K}_{Y}(\lambda^{K}, \lambda^{\bar{K}})\geq0$.  
The monotonicity of the $\rho^{K}_{Y}$ implies that the social welfare ($Y=W$) and the platform's revenue ($Y=R$) are maximized by matching all sellers to all buyers.
Hence, the market is complete-matched on both sides.

Next, consider $\rho^{\bar{K}}_{Y}(\underline{\lambda}^{\bar{K}}, \underline{\lambda}^{K})<0\leq \rho^{K}_{Y}(\underline{\lambda}^{K}, \underline{\lambda}^{\bar{K}})$.
Suppose $\eta^{K}_{Y}(\underline{\lambda}^{K}, \underline{\lambda}^{\bar{K}})\geq 0$, then the platform uses a matching rule $\tau^{K}_{\Gamma}(\lambda^{K})\geq \underline{v}^{\bar{K}}$ for some $\lambda^{K}\in \Lambda^{K}$. 
Let $\lambda^{K}_{a}$ be any type in $\Lambda^{K}$, such that $\tau^{K}$ is strictly decreasing at the right neighborhood of $\lambda^{K}_{a}$.
Hence, the marginal effect of reducing $\tau^{K}(\lambda^{K})$ below $\lambda^{\bar{K}}$ is $\eta^{K}_{Y}(\lambda^{K}, \lambda^{\bar{K}})$.
Let $[\lambda^{K}_{\ell}, \lambda^{K}_{h}]\subset \Lambda^{K}$, in which $\tau^{K}(\lambda^{K}) = \lambda^{\bar{K}}$ for all $\lambda^{K}\in [\lambda^{K}_{\ell}, \lambda^{K}_{h}]$, i.e., $\tau^{K}$ specifies a constant cut-off type $\lambda^{\bar{K}}$, in which the marginal effect of the reduction of $\tau^{K}(\lambda^{K})$ below $\lambda^{\bar{K}}$ is $\int^{\lambda^{K}_{h}}_{\lambda^{K}_{\ell}}\eta^{K}_{Y}(x, \lambda^{\bar{K}})dx$.
Under Assumption \ref{assp:regularity}, we have $\eta^{K}_{\Gamma}(\lambda^{K}, \lambda^{\bar{K}})>0$ for any $\lambda^{K}\in \Lambda^{K}$ and $\lambda^{\bar{K}}\in \Lambda^{\bar{K}}$, which implies that a complete-matched pattern for both sides is optimal.
Let $g^{K}_{Y} = \inf\{\lambda^{K}\in\Lambda^{K}: \rho^{K}_{Y} \geq 0\}$.

%Suppose $f^{K}(\lambda^{K})>0$ for all $\lambda^{K}\in\Lambda^{K}$.
The platform's objective functions can be written in terms of $U^{K}_{Y}$ as follows: for $Y\in\{W,R\}$,
$$
\begin{aligned}
Z_{Y}(\{\bm{\tau}, \bm{\delta}\}) =& \sum_{K\in{S,B}}\int_{\delta^{K}}^{\bar{\lambda}^{K}}U^{K}_{Y}(x, \tau^{K}(x))f^{K}(x) dx\\
% =& \sum_{K\in{S,B}}\int_{\delta^{K}}^{\bar{\lambda}^{K}} \theta^{K}_{Y}(x, \tau^{K}(x))dx.
\end{aligned}
$$

Since $\eta^{K}_{Y}(\underline{\lambda}^{K}, \underline{\lambda}^{\bar{K}})<0$, $g^{K}_{Y}>\underline{\lambda}^{K}$.
From Appendix \ref{app:prop_uniqueness}, $\delta^{K}\in [\underline(\lambda)^{K}, g^{K}_{Y} ]$ at the optimum.
Moreover, $\delta^{\bar{K}}_{Y}\in[\underline{\lambda}^{\bar{K}}, g^{K}_{Y}]$ and $\tau^{K}(g^{K}_{Y}) = g^{\bar{K}}_{Y}$.
Assume that $g^{K}_{Y}>\underline{\lambda}^{\bar{K}}$.
(If not, then $\delta^{\bar{K}} = \underline{\lambda}^{\bar{K}}$  and $\tau^{K}_{K}(\lambda^{K}) = \underline{\lambda}^{K}$ for all $v_{k}\geq g^{K}_{Y}$).

Hence, we can partition $Z_{Y}$ into two independent problem, for $K\in{S,B}$, 
$$
\begin{aligned}
Z^{K}_{Y} = \int^{g^{K}_{Y}}_{\delta^{K}} U^{K}_{Y}(x,\tau^{K}(x))f^{K}(x) dx + int_{g^{\bar{K}}_{Y}}^{\bar{\lambda}^{\bar{K}}} U^{\bar{K}}_{Y}(x,\tau^{\bar{K}}(x))f^bar{K}(x) dx.
\end{aligned}
$$
Here, the problem $Z^{K}_{Y}$ can be solved piecewisely.
Let $\lambda^{K}_{\Delta}=\inf\{\lambda^{K}\in[\underline{\lambda}^{K}, g^{K}_{Y}]: \eta^{K}_{Y}(\lambda^{K}, \kappa(\lambda^{K})) = 0\}$.
Then, the following cut-off function maximizes $Z^{K}_{Y}$:
$$
\tilde{\tau}^{K}(\lambda^{K})=\begin{cases}
\bar{\lambda}^{\bar{K}} & \text{ if, } \lambda^{K}\in[\underline{\lambda}^{K},  \lambda^{K}_{\Delta}],\\
\kappa(\lambda^{K}) & \text{ if, } \lambda^{K}\in (\lambda^{K}_{\Delta}, \bar{\lambda}^{K}].
\end{cases}
$$

Next, let cut-off function associated to the matching rule  $\tau^{K}=\tilde{\tau}^{K}$ for any $\lambda^{K}\in [ \lambda^{K}_{\Delta}, g^{K}_{Y}]$.
Given $\tau^{K}: [\delta^{K}, g^{K}_{Y}] \mapsto [\delta^{\bar{K}}, g^{\bar{K}}_{Y}]$, reciprocally, we have $\tau^{\bar{K}}:[\delta^{\bar{K}}, g^{\bar{K}}_{Y}] \mapsto  [\delta^{K}, g^{K}_{Y}] $.
It is straightforward to see that matching rule characterized by the above $\tau^{K}$ with $\delta^{K}$ is unique in the sense that it specifies a unique cut-off type for each matched type on side $K$.
It is easy to see that the market if bottom-eliminated on side $K$ if $\lambda^{K}_{\Delta}>\underline{\lambda}^{K}$ and top-reserved on side $\bar{K}$ if $\lambda^{K}_{\Delta} = \underline{\lambda}^{K}$.
From Assumption \ref{assp:regularity}, we have $\eta^{K}_{Y}(\underline{\lambda}^{K}, \bar{\lambda}^{K}) <0$.

Next consider that $\lambda^{K}_{\Delta} = \underline{\lambda}^{K}$, in which there exists a $\kappa(\underline{\lambda}^{K})\in[g^{\bar{K}}_{Y},\bar{\lambda}^{K} ]$ such that $\eta^{K}_{Y}(\underline{\lambda}^{K},\kappa(\underline{\lambda}^{K}))=0$.
Suppose $\kappa(\underline{\lambda}^{K})<\bar{\lambda}^{\bar{K}}$. 
From Assumption \ref{assp:regularity}, $\eta^{K}_{Y}(\underline{\lambda}^{K}, \bar{\lambda}^{\bar{K}})>0$.
Hence, we have that if $\eta^{K}_{Y}(\underline{\lambda}^{K}, \bar{\lambda}^{\bar{K}})>0$, the market is complete-matched on side $\bar{K}$ and top-reserved on side $K$.
Finally, consider $\kappa(\underline{\lambda}^{K}) = \bar{\lambda}^{\bar{K}}$, in which $\delta^{K} = \underline{\lambda}^{K}$ and $\tau^{K}(\underline{\lambda}^{K}) = \bar{\lambda}^{\bar{K}}$, i.e., the market is not top-reserved on side $\bar{K}$ and complete-matched on side $K$.

\end{document}